\documentclass[%
preprint,
superscriptaddress,
 amsmath,amssymb,
 aps,
]{revtex4-1}

\usepackage{graphicx}
\usepackage{dcolumn}
\usepackage{bm}
\usepackage{amssymb}
\usepackage{caption}
\usepackage{subcaption}
\usepackage{amsmath}
\usepackage[T1]{fontenc}
\usepackage[utf8]{inputenc}
 \usepackage{setspace}
 \usepackage{float}
 \usepackage{array}
 \usepackage{multirow}
\newcolumntype{P}[1]{>{\centering\arraybackslash}p{#1}}
\newcolumntype{M}[1]{>{\centering\arraybackslash}m{#1}}

\begin{document}


\title{Dynamics of a droplet migration in oscillatory and pulsating microchannel flows and prediction and uncertainty quantification of its lateral equilibrium position using Multi Fidelity Gaussian processes}

\author{Ali Lafzi}
 \affiliation{Department of Agricultural and Biological Engineering, Purdue University, West Lafayette, Indiana 47907, USA}

\author{Sadegh Dabiri}
 \affiliation{Department of Agricultural and Biological Engineering, Purdue University, West Lafayette, Indiana 47907, USA}




\begin{abstract}
Dynamics of a droplet in oscillatory and pulsating flows of a Newtonian fluid in a microchannel has been studied numerically. The effects of oscillation frequency, surface tension, and channel flow rate have been explored by simulating the drop within a microchannel. These types of flows introduce new equilibrium positions for the drop compared to steady flows with similar conditions. The simulation results are very sensitive to the grid resolution due to the unsteady behavior of the base flow. Therefore, a set of fine grids have been used in this study to capture the physics of this problem more accurately. However, these fine grids make the computations significantly expensive. Therefore, a Multi Fidelity Gaussian processes method with two levels of fidelity has been used to predict the results of the remaining fine-grid simulations along with their uncertainties based on their correlations with those of the coarse-grid cases over a wide range of input parameters. 

\end{abstract}

\maketitle


\section{\label{sec1}Introduction}
Transient dynamics of droplets and their lateral equilibrium positions in microchannels depend on several parameters including the rheology of the carrier fluid, the geometry of channel, the channel flow rate, and the drop characteristics such as shape, size, interfacial tension, etc, which can be deployed for numerous biological and biomedical applications \cite{gossett2010label, toner2005blood, gossett2012inertial, karimi2013hydrodynamic, di2007continuous}.  

Most of the effective parameters underlying the physics of particle motion in microchannels and label-free approaches are inherent to the system, particle properties, and are out of human's complete control. Therefore, adding an extra parameter as a tool for more direct control over the dynamics of particles and better manipulating them can be very beneficial; especially, coming up with equilibrium positions between the channel center and the wall can be very useful for separation purposes \cite{marson2018inertio}. The other restriction in inertial microfluidics is the limitation of working with particle sizes of a few microns or larger only \cite{mutlu2018oscillatory}. This is because of the inverse relationship between the particle size and its traveling distance until focusing, which means that the separation of smaller particles requires longer channels \cite{lan2012numerical}. 

Exploiting an oscillatory flow in the channel has been recommended as a solution to address the aforementioned issues \cite{mutlu2018oscillatory, lafzi2020inertial}. The oscillation frequency as the new extra parameter does not have the disadvantage of adversely affecting the biological properties of cells like most of the active methods \cite{erlandsson2011electrolysis}. Besides, this type of flow as an alternative to the traditional microfluidics has attracted researchers in recent years. \citeauthor{chaudhury2016droplet} \cite{chaudhury2016droplet} have observed a complicated trajectory of the droplet in the lateral direction suspended in an oscillatory flow as opposed to the smooth cross-stream migration under steady flow conditions. \citeauthor{pawlowska2017lateral} \cite{pawlowska2017lateral} have reported the use of hydrogel nanofilaments as an appropriate substitute for the long and deformable macromolecules and studied their dynamics in an oscillatory microchannel flow. They have shown that the final position of these particles fluctuates around the flow axis \cite{pawlowska2018diffusion}. \citeauthor{sarkar2000deformation} \cite{sarkar2000deformation} have investigated the dynamics of a viscoelastic drop in steady and oscillating extensional flows and have shown that although their deformation behaviors are naturally dissimilar, their maximum values are close in the long-time limit.

The presented study investigates different aspects of the droplet migration in oscillatory and pulsating microchannel flows numerically and compares them with those in the steady ones. The main challenge of performing this numerical work is using a sufficient grid resolution to resolve the underlying physics more accurately. Since the background flow is unsteady, the mesh has to be fine enough to capture the velocity gradients and lift forces acting on the drop correctly. Consequently, the transient dynamics of the drop and its final equilibrium position, which is the main goal of the presented study, are all affected by this choice. Therefore, after carrying out the mesh dependence study, a very fine grid has been selected to enable getting the realistic results we are interested in. However, running simulations with these grids can take a few months in some cases. Hence, it is impossible to get all the data in the wide range of input parameters we are looking for by just running all these simulations. An alternative approach is to produce some new data using a coarser grid resolution and implement the Multi Fidelity Gaussian processes (MFGP) algorithm to predict the outputs of the finer mesh. \citeauthor{perdikaris2017nonlinear} \cite{perdikaris2017nonlinear} have proposed a probabilistic framework and a recursive version of MFGP. This method has been tested in several benchmark
problems involving both synthetic and real multi-fidelity datasets such as the one employed by \citeauthor{babaee2016multi} \cite{babaee2016multi}. The implementation of the recursive MFGP and the obtained results in this study are elaborated in the following sections.
\section{Methodology\protect\\}
\subsection{The fluid dynamics simulations}
\begin{figure}[h]
  \centering
  \includegraphics[height=2.4in,width=3.1in]{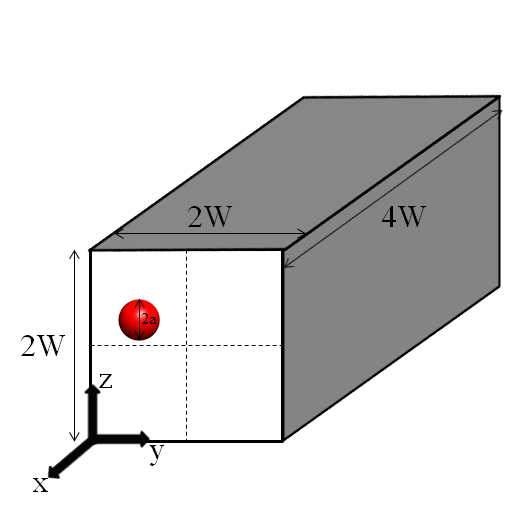}
   \caption{Schematic of the problem setup}
   \label{geometry}
\end{figure}

A single Newtonian droplet has been placed in a laminar flow of an incompressible Newtonian fluid in a rectangular microchannel with a square cross section. A schematic of the configuration is illustrated in Fig. \ref{geometry}. The density and viscosity ratios ($\eta$ and $\lambda$, respectively) are set to 1 for most of the simulations. The front-tracking method \cite{unverdi1992front} is used to update the interface position. In this method, the main governing equations are solved in the fixed Eulerian grid, and the obtained information is used to update the properties across the droplet surface containing thousands of moving Lagrangian elements. The governing equations to be solved in the entire computational domain are the following:
\begin{equation}
\nabla \cdot \textbf{u}=0,
\end{equation}
\begin{equation}\label{eq:1}
\frac{\partial (\rho \textbf{u})}{\partial t}+\nabla \cdot (\rho \textbf{uu})=-\nabla p+\nabla \cdot \boldsymbol{\tau} +\iint{} \gamma\kappa\delta(\boldsymbol{x}-\boldsymbol{x_i}) \textbf{n} dA,
\end{equation}
where ${\rho}$ is the density of the fluid, $p$ represents the pressure, \textbf{u} is the velocity vector, $t$ is the time, $\boldsymbol{\tau}=\mu(\nabla \textbf{u}+\nabla \textbf{u}^T)$ is the stress tensor in which $\mu$ is the fluid viscosity, $\kappa$ is the curvature at the interface, $\gamma$ is the interfacial tension, $\delta$ is the Dirac delta function, $\mbox{\boldmath$x$}$ is an arbitrary location in the whole computational domain, $\mbox{\boldmath$x_i$}$ is such position on the drop interface, and \textbf{n} is the unit normal vector to a point on the interface. The given delta function is defined as:
\begin{equation}\label{eq:9}
\delta(x)=\tilde{D}(x)\tilde{D}(y)\tilde{D}(z),
\end{equation}
\begin{equation}
\tilde{D}(x)=\frac{1}{4\Delta}(1+\cos(\frac{\pi}{2\Delta}(x))) ,  |x|\le 2\Delta,
\end{equation}
where $\Delta$ is the constant Eulerian grid size.

To generate the oscillatory flow, a cosine wave of pressure gradient with a constant amplitude (in the form of ${P_0}cos({\omega}t)$) is applied along the channel ($x$ direction) to change the direction of the flow symmetrically. For the pulsating case, this pressure gradient has the shape of ${P_0}(a+bcos({\omega}t))$, where $a$ and $b$ are the weights of the steady and oscillatory components, respectively, and $a+b=1$. The periodic boundary condition is applied in the $x$ direction, and the no-slip condition is applied on the walls in the $y$ and $z$ directions. $W$ and $U_0$ (channel centerline velocity corresponding to the steady case) are used as the characteristic length and velocity, respectively. In other words, $x^*={\frac{x}{W}}$, $u^*={\frac{u}{U_0}}$, $t^*={\frac{t}{{\frac{W}{U_0}}}}$, $P^*={\frac{P}{\mu{\frac{U_0}{W}}}}$, $T^*={\frac{T}{{\frac{W}{U_0}}}}$ (where $T$ is the period), and $\omega^*={\frac{2\pi}{T^*}}$. Three dimensionless parameters describe and affect the motion of the drop: (i) Reynolds number $Re={\frac{{\rho}U_02W}{\mu}}$, expressing the ratio between inertial forces to viscous ones (ii) Capillary number $Ca={\frac{\mu U_0}{\gamma}}$, which denotes the ratio
of viscous stress to the interfacial tension, where high $Ca$ corresponds to a highly deformable drop (iii) The dimensionless oscillation frequency ($\omega^*$). The effect of the frequency can also be embedded in the Womersley number $Wo=(\frac{\omega W^2}{\nu})^\frac{1}{2}$, where $\nu$ is the kinematic viscosity of the fluid. This number compares the transient inertial effects to viscous forces \cite{dincau2020pulsatile}. The blockage ratio of the drop (${\frac{a}{W}}$) is constant and equals to $0.3$ for most of the cases studied here. The drop is assumed to have a spherical initial shape and is released at ${\frac{y}{W}}=0.55$ and ${\frac{z}{W}}=1.07$. The initial location of the drop is arbitrary since it does not alter its equilibrium position \cite{pan2016motion, lan2012numerical, chaudhury2016droplet, razi2017direct}. The axes of symmetry have been avoided. A fine grid of $196\times 114 \times 114$ (high fidelity level) and a coarse grid of $128\times 76 \times 76$ (low fidelity level) in the $x$, $y$, and $z$ directions, respectively, is used for the simulations having $Re=10$. The simulations with higher $Re$ numbers require even a finer grid resolution ($256\times 152 \times 152$). Since this adds another level of fidelity to the problem, we have omitted the effect of $Re$ in the second part of the paper for more simplicity. Therefore, the input parameters that affect the output of interest, which is the distance of the equilibrium position of the drop from the channel center, are reduced to the Capillary number ($Ca$) and frequency ($\omega^*$).
\subsection{Multi fidelity GP}
Gaussian processes (GP) is an example of a continuous stochastic process in which, a multi-dimensional Gaussian distribution is assigned to the function values at the input points as the prior knowledge. In other words, the value of the function that we are aiming to predict follows a Gaussian probability density at each point:
$$
\mathbf{f}_{1:n} | \mathbf{x}_{1:n} \sim \mathcal{N}\left(\mathbf{m}(\mathbf{x}_{1:n}), \mathbf{K}(\mathbf{x}_{1:n}, \mathbf{x}_{1:n}) \right),
$$
$$
\mathbf{m}(\mathbf{x}_{1:n}) =
\left(
m(\mathbf{x}_1),
\dots,
m(\mathbf{x}_n)
\right),
$$
$$
\mathbf{K}(\mathbf{x}_{1:n}, \mathbf{x}_{1:n}) = \left(
\begin{array}{ccc}
k(\mathbf{x}_1,\mathbf{x}_1) & \dots & k(\mathbf{x}_1, \mathbf{x}_n)\\
\vdots & \ddots & \vdots\\
k(\mathbf{x}_n, \mathbf{x}_1) & \dots & k(\mathbf{x}_n, \mathbf{x}_n)
\end{array}
\right).
$$
Where $\mathbf{m}$ is the mean vector, and $\mathbf{K}$ is the covariance matrix whose elements denote the correlations between the input points. These correlations are quantified by a covariance (kernel) function that encodes our prior belief and knowledge about the targeted function like its continuity, differentiability, etc. The choice of this function is very important for a good prediction; otherwise, we need to have sufficient amount of data to compensate this lack of knowledge. One of the most typical examples for this function is the squared exponential (SE) kernel given below:
$$
k(\mathbf{x}, \mathbf{x}') = v\exp\left\{-\frac{1}{2}\sum_{i=1}^d\frac{(x_i - x_i')^2}{\ell_i^2}\right\},
$$
In which, $v$ is known as the signal strength, and $\ell_i$ is the length scale of the $i$-th input dimension of the GP. In this study, we also use the SE function as we believe the function of interest is continuous and infinitely differentiable.

If we have $n$ observations (training set), consisting of $\mathbf{x}_{1:n}=(\mathbf{x}_1,\dots,\mathbf{x}_n)$ and $\mathbf{y}_{1:n}=(y_1,\dots,y_n)$, and $n^*$ test points ($\mathbf{x}^*_{1:n^*}$) that we would like to predict the function values at ($\mathbf{f}^*_{1:n^*}$), we use the following model according to the definition of GP as our likelihood:
$$
p(\mathbf{f}_{1:n}, \mathbf{f}^*_{1:n^*} | \mathbf{x}_{1:n}, \mathbf{x}^*_{1:n^*}) = \mathcal{N}\left(
\left(
\begin{array}{c}
\mathbf{f}_{1:n}\\
\mathbf{f}^*_{1:n^*}
\end{array}
\right)\middle |
\left(
\begin{array}{c}
\mathbf{m}(\mathbf{x}_{1:n})\\
\mathbf{m}(\mathbf{x}^*_{1:n^*})
\end{array}
\right),
\left(
\begin{array}{cc}
\mathbf{K}(\mathbf{x}_{1:n}, \mathbf{x}_{1:n}) & \mathbf{K}(\mathbf{x}_{1:n}, \mathbf{x}^*_{1:n^*})\\
\mathbf{K}(\mathbf{x}^*_{1:n^*}, \mathbf{x}_{1:n}) & \mathbf{K}(\mathbf{x}^*_{1:n^*}, \mathbf{x}_{1:n^*})
\end{array}
\right)
\right),
$$
After performing the Bayes' rule and assuming the observations are not noisy ($\mathbf{y}_{1:n}=\mathbf{f}_{1:n}$) and $\mathcal{D} = (\mathbf{x}_{1:n}, \mathbf{y}_{1:n})$ being the observed data, we get the posterior distribution for $\mathbf{f}^*$ as below:
$$
p(\mathbf{f}^*_{1:n^*}| \mathbf{x}^*_{1:n^*}, \mathcal{D}) = \mathcal{N}\left(\mathbf{f}^*_{1:n^*}\middle| \mathbf{m}_n(\mathbf{x}^*_{1:n^*}), \mathbf{K}_n(\mathbf{x}^*_{1:n^*},\mathbf{x}^*_{1:n^*})\right),
$$
Where the posterior mean function is:
$$
m_n(x) = m(x) + \mathbf{k}(x,\mathbf{x}_{1:n})\left(\mathbf{K}(\mathbf{x}_{1:n},\mathbf{x}_{1:n})\right)^{-1}\left(\mathbf{y}_{1:n} - \mathbf{m}(\mathbf{x}_{1:n})\right),
$$
And the posterior covariance function is:
$$
k_n(x, x') = k(x,x') - \mathbf{k}(x,\mathbf{x}_{1:n})\left(\mathbf{K}(\mathbf{x}_{1:n},\mathbf{x}_{1:n})\right)^{-1}\mathbf{k}^T(x,\mathbf{x}_{1:n}),
$$
With $\mathbf{k}(x,\mathbf{x}_{1:n}) = \left(k(x,\mathbf{x}_1),\dots,k(x,\mathbf{x}_n)\right)$ being the cross-covariance vector.

Multi-fidelity modelling enables accurate inference
of quantities of interest by synergistically combining
realizations of low-cost/low-fidelity models with
a small set of high-fidelity observations. This is
particularly effective when the low and high-fidelity
models exhibit strong correlations and can lead to
significant computational gains over approaches
that solely rely on high-fidelity models \cite{perdikaris2017nonlinear}. The MFGP method introduced by \citeauthor{perdikaris2017nonlinear} \cite{perdikaris2017nonlinear} is fundamentally very similar to that of the \citeauthor{kennedy2000predicting} \cite{kennedy2000predicting}, but it is also capable of capturing the more complex nonlinear cross-correlations between the data. Assuming that we have $s$ fidelity levels (which is 2 in our case), based on this method, the GP has to be done $s$ times recursively. In other words, the GP is done in its standard, usual way on the first level, and for the next levels, the outputs of the previous level $t-1$ are the inputs for the level $t$. This is done in the way described below \cite{perdikaris2017nonlinear}:
$$f_t(\mathbf{x}) = g_t(\mathbf{x}, f_{*t-1} (\mathbf{x})),$$
Where $f_{*t-1} (\mathbf{x})$ is the posterior distribution of the previous level $t-1$ evaluated at the $\mathbf{x}$, being the inputs of the current level $t$. The unknown function $g_t$ follows the description of GP:
$$g_t \sim \mathcal{G}\mathcal{P}( f_t|0, k_t((\mathbf{x}, f_{*t-1} (\mathbf{x})), (\mathbf{x}', f_{*t-1} (\mathbf{x}')); \theta_t))$$
According to \citeauthor{le2014recursive} \cite{le2014recursive}, this scheme has the same posterior distribution predicted by the fully coupled scheme of \citeauthor{kennedy2000predicting} \cite{kennedy2000predicting}. This procedure implies two conditions: (i) The training sets need to have a nested structure (i.e. $\mathbf{x}_t\subseteq \mathbf{x}_{t-1}$) (ii) The Markov property:
$$cov\left\{f_t(\mathbf{x}), f_{t-1}(\mathbf{x}') | f_{t-1}(\mathbf{x})\right\} = 0, \forall \mathbf{x} \ne \mathbf{x}'$$
which translates into assuming that given the nearest point $f_{t-1}(\mathbf{x})$, we can learn nothing more
about $f_t(\mathbf{x})$ from any other model output $f_{t-1}(\mathbf{x}')$, for any $\mathbf{x} \ne \mathbf{x}'$.

Since $f_{*t-1}$ and $\mathbf{x}$ belong to inherently 2 different spaces, a more structured kernel function, coupling the elements from the same space together, is the following \cite{perdikaris2017nonlinear}:
$$k_{t_g} = k_{t_\rho} (\mathbf{x}, \mathbf{x}'; \theta_{t_\rho} ) \cdot k_{t_f} ( f_{*t-1} (\mathbf{x}), f_{*t-1} (\mathbf{x}'); \theta_{t_f} ) + k_{t_\delta} (\mathbf{x}, \mathbf{x}'; \theta_{t_\delta} ),$$
where $k_{t_\rho}$, $k_{t_f}$, and $k_{t_\delta}$ are valid covariance functions, and $\theta_{t_\rho}$, $\theta_{t_f}$, and $\theta_{t_\delta}$  are their hyperparameters, respectively. For our application, we have chosen the SE kernel function:
$$
k(x, x'; \theta_t) = \sigma_t^2\exp\left\{-\frac{1}{2}\sum_{i=1}^d\mathcal{W}_{i,t}(x_i - x_i')^2\right\},
$$
where $\sigma_t^2$ is a variance parameter, and $\bigl\{\mathcal{W}_{i,t}\bigr\}_{i=1}^d$ are the Automatic Relevance Determination (ARD) weights corresponding to the fidelity level $t$.

As previously discussed, the posterior distribution of the first level is obtained by the normal GP. Therefore, it is Gaussian. However, this is not generally the case for the subsequent levels, except for the case of $t=2$.  Therefore, for all the cases with
$t \geq 2$, we have to perform predictions given uncertain inputs, where the uncertainty is propagated along each recursive step. Thus, the posterior distribution is given by \cite{perdikaris2017nonlinear}:
$$p( f_{*t} ( \mathbf{x}_{*})) :=p( f_{t}(\mathbf{x}_{*}, f_{*t-1} (\mathbf{x}_{*}))|f_{*t-1} , \mathbf{x}_{*}, \mathbf{y}_t, \mathbf{x}_t)
=\int p( f_{t}(\mathbf{x}_{*}, f_{*t-1} (\mathbf{x}_{*}))|\mathbf{y}_t, \mathbf{x}_t, \mathbf{x}_{*})p(f_{*t-1} (\mathbf{x}_{*})) d\mathbf{x}_{*}$$
The predictive mean and variance of all posteriors $p( f_{*t} ( \mathbf{x}_{*})), t \geq 2$ is calculated using Monte Carlo integration of this equation. We can then sample from this distribution.
\section{Results and Discussion\protect\\}
\subsection{\label{sec:level2}Oscillatory flow}
Dynamics of the droplet in an oscillatory flow in the microchannel has been studied, and the effects of $Re$, $Ca$, and $\omega^*$ have been investigated. $Re$ ranges between $10$ and $100$, $Ca$ ranges between $0.09$ and $10$, and  $\omega^*$ values are chosen such that for a channel with a square cross-section of $100 \mu m$ and water as the working fluid at room temperature, the frequency values range between 2Hz and 1600Hz (or a corresponding $Wo$ number between $0.2$ and $6.3$). Although a maximum frequency of 200Hz is mostly reported in the literature \cite{dincau2020pulsatile}, recent works have claimed of generating frequencies of around 1KHz \cite{vishwanathan2020generation}. The equilibrium position of the drop is a result of the competition between the lateral lift forces acting on it, including the wall effect and the deformation-induced lift forces, both acting towards the channel center, and the shear gradient force acting towards the wall \cite{fay2016cellular, zhang2016fundamentals, rivero2018bubble}. Magnus and Saffman lift forces are often very small compared to the other mentioned components and can be neglected  \cite{zhang2016fundamentals, martel2014inertial, stoecklein2018nonlinear}. The boundary wall causes the particle to have rotational and translational velocities different from those of the adjacent fluid, which is caused by an uneven distribution of vorticities around the particle \cite{ho1974inertial, bagchi2002shear}. This induces a higher pressure in the gap between the particle and the wall, which repels the particle away from the wall \cite{zhang2016fundamentals}. The existing curvature of the fluid velocity profile makes the magnitude of the velocity of the fluid on the wall side much higher than the channel center side from the particle frame of reference.  This inequality causes a low pressure on the wall side leading to a shear gradient lift force that pushes the particle towards the wall \cite{zhang2016fundamentals}. Following the analytical results of \citeauthor{chan1979motion} \cite{chan1979motion}, the deformability-induced lift force for droplets or bubbles that have a distance higher than their diameter from the wall, which is the case in our simulations, is given by \cite{stan2013magnitude}:
\begin{equation}\label{deformation force}
F_{L,deformation} =Ca_p\mu V_{avg}a(\frac{a}{W})^2\frac{d}{W}f(\lambda),
\end{equation}
\begin{equation}\label{eq:3}
f(\lambda) =\frac{128\pi}{(\lambda+1)^3}\left\{\frac{11\lambda+10}{140}(3\lambda^2-\lambda+8)-\frac{3(19\lambda+16)}{14(3\lambda+2)}(2\lambda^2+\lambda-1)\right\},
\end{equation}
Where $Ca_p={\frac{\mu U_0}{\gamma}}{\frac{a}{W}}$ is the drop capillary number, $V_{avg}$ is the average velocity of the carrier fluid across the channel, $d$ is the distance of the drop from the channel center, and $\lambda$ is the viscosity ratio between the inner and outer fluids. 

The code has been validated by comparing the drop deformations at $Ca=0.2$ and different Deborah numbers with those of the \citeauthor{aggarwal2007deformation} \cite{aggarwal2007deformation}. The results are in good agreement with a maximum error of $0.72 \%$. To validate the inertial effects, the focal points of the drop at $Re=8.25$, $Ca=0.18$, $\frac{a}{W}=0.2$ and $Re=21$, $Ca=0.14$, $\frac{a}{W}=0.3$ have been compared with those presented by \citeauthor{marson2018inertio} \cite{marson2018inertio}. Our obtained focal points lie within their corresponding uncertainty bands. Furthermore, we have shown that the numerical results are independent of the distance between $2$ consecutive drops in an infinite domain in the flow direction. This has been done by comparing the drop trajectory at $Re=10$, $Ca=1$, and $\omega^*=0.1$ for three different channel lengths of $4W$, $6W$, and $8W$ in our simulation setup. The maximum difference between the drop trajectories for $L=4W$ and $L=6W$ is $0.0003W$, and the one between those of $L=4W$ and $L=8W$ is $0.0005W$. The results have also been shown to be grid independent, by comparing the equilibrium positions for the case of $Re=10$, $Ca=1$, and $\omega^*=0.1$ with two different grids of $196\times 114 \times 114$ and $256\times 152 \times 152$. The difference between their focal distances from the center is $0.0009W$.

\begin{figure}[ht]
  \centering
  \includegraphics[width=4.3in]{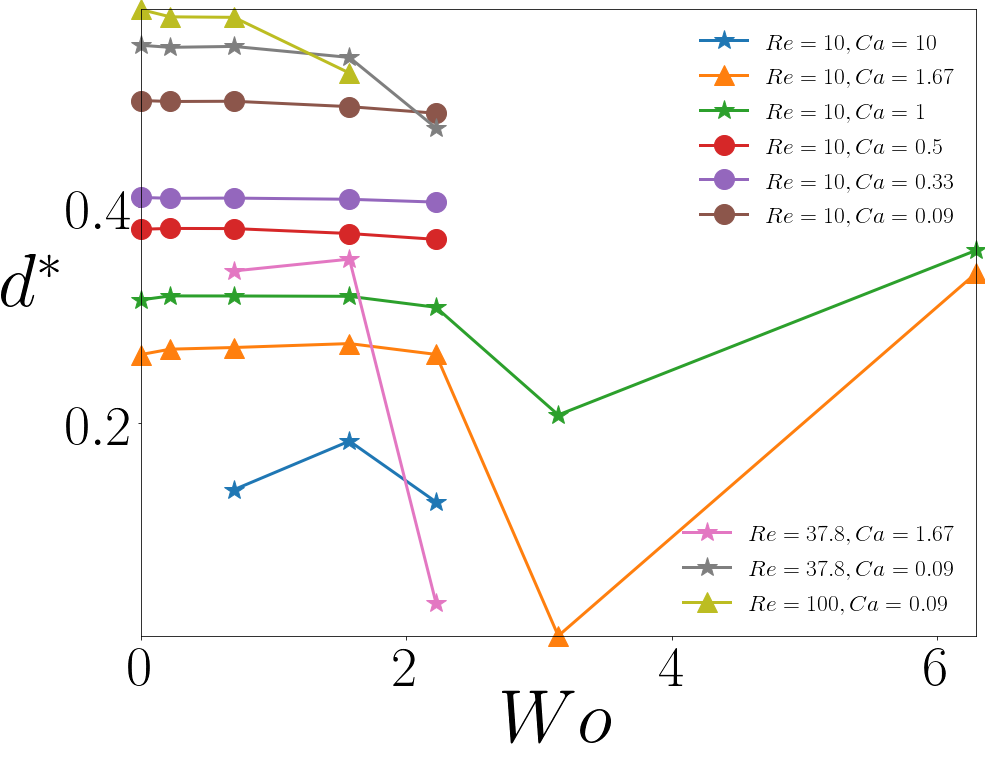}
   \caption{Distance of the equilibrium position of the droplet from the channel center as a function of $Wo$ for different cases}
   \label{focal point}
\end{figure}

Figure \ref{focal point} shows the distance of the droplet focal point from the channel center ($d^*$) at different values of $Wo$, $Ca$, and $Re$. The drop focal point in the steady flow moves towards the center by increasing the $Ca$ due to the increase in the deformation force \cite{pan2016motion, lan2012numerical, hadikhani2018inertial} and shifts towards the wall as $Re$ increases because of the improvement in the strength of the shear gradient force \cite{mortazavi2000numerical, hadikhani2018inertial, di2009particle}. Since $W$ and $\nu$ are constant in this work, $\omega^*$ and $Wo$ are directly related to each other. Hence, we can use them interchangeably. The droplet travels at locations far from the wall; so the wall lift can be neglected in our study \cite{zhang2016fundamentals, ho1974inertial}. Both deformation and shear gradient lift forces, as the remaining active forces, depend on $V_{avg}$ and $d^*$ that vary as the simulations proceed. The dependence of deformation force on these 2 parameters is apparent from equation \ref{deformation force}. The parameter $d^*$ and the flow velocity at the drop location affect the magnitude of the difference between the velocities on the wall and center sides from the drop frame of reference. Hence, both $d^*$ and $V_{avg}$ determine the magnitude of the shear gradient force. For non-steady flows, including oscillatory and pulsating ones, $V_{avg}$ is time-dependent. The average of this $V_{avg}$ in each corresponding periodic cycle decreases as the $\omega^*$ (or $Wo$) increases. Consequently, the averages of both forces in one periodic cycle change by changing the $Wo$ value keeping other parameters fixed, leading to different equilibrium positions as we can see in Fig. \ref{focal point}. A complete explanation of the relationship between the focal point and parameters like $\omega^*$, $Ca$, and $Re$ can be found in our previous work \cite{lafzi2020inertial}. According to this figure, the focal point is closest to the channel center at the highest $Wo$ except for $Re=10$ and $Ca=1$ and $Re=10$ and $Ca=1.67$. This can be explained based on the shape of the flow velocity profile elaborated below.

\begin{figure}
\centering
  \begin{subfigure}[ht]{.485\textwidth}
  \centering
  \includegraphics[height=.34\textheight]{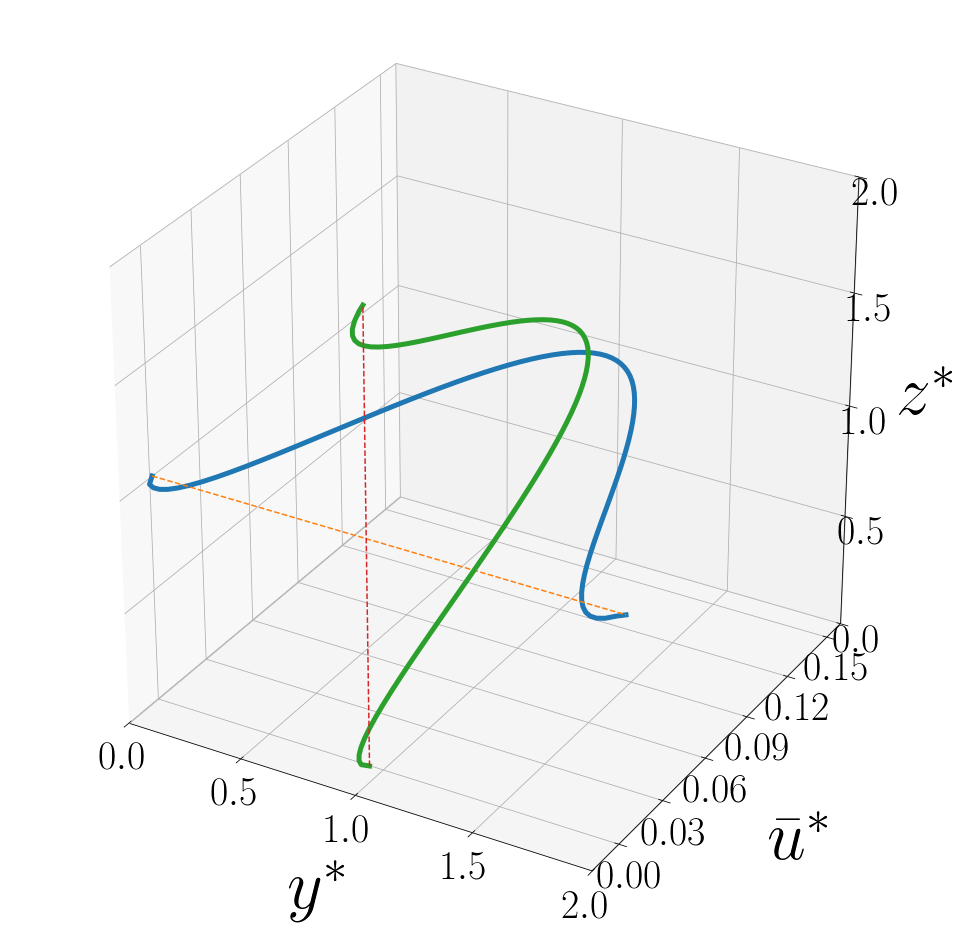}
  \caption{}
  \label{velocity_omega=2}
  \end{subfigure}
  ~
  \begin{subfigure}[ht]{.485\textwidth}
  \centering
  \includegraphics[height=.34\textheight]{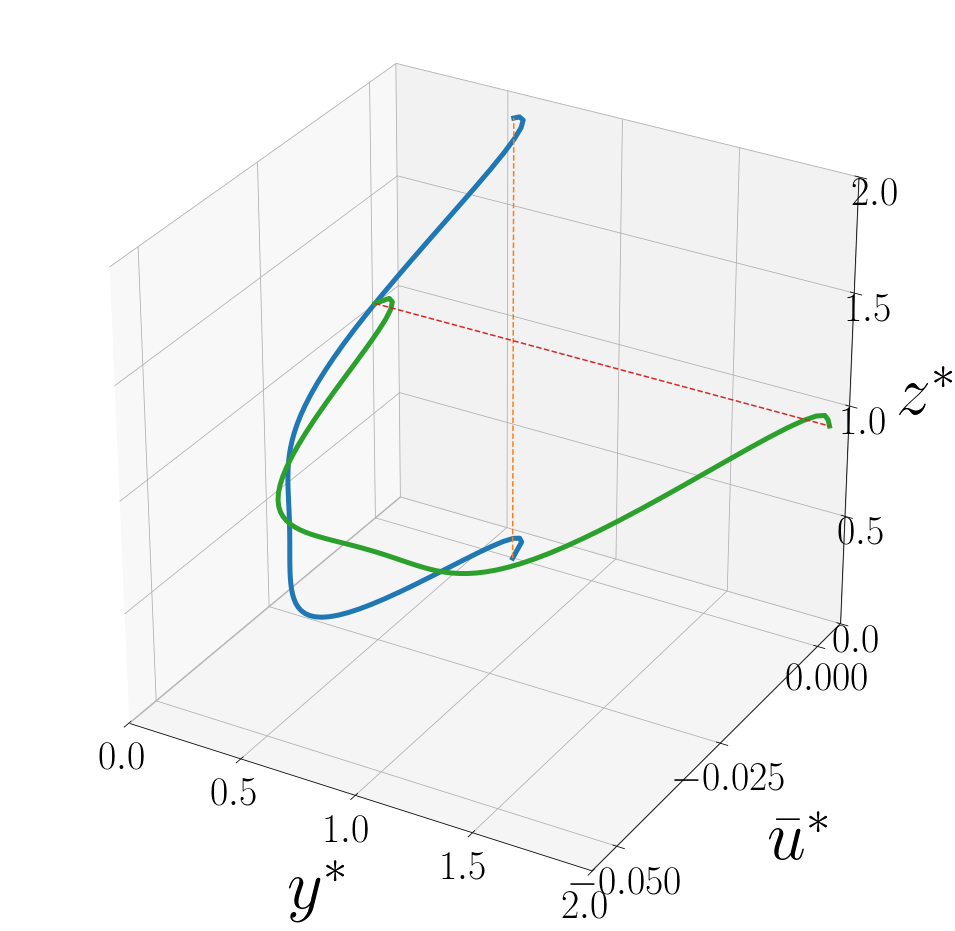}
  \caption{}
  \label{velocity_omega=8}
  \end{subfigure}
  ~
  \begin{subfigure}[ht]{.485\textwidth}
  \centering
  \includegraphics[height=.34\textheight]{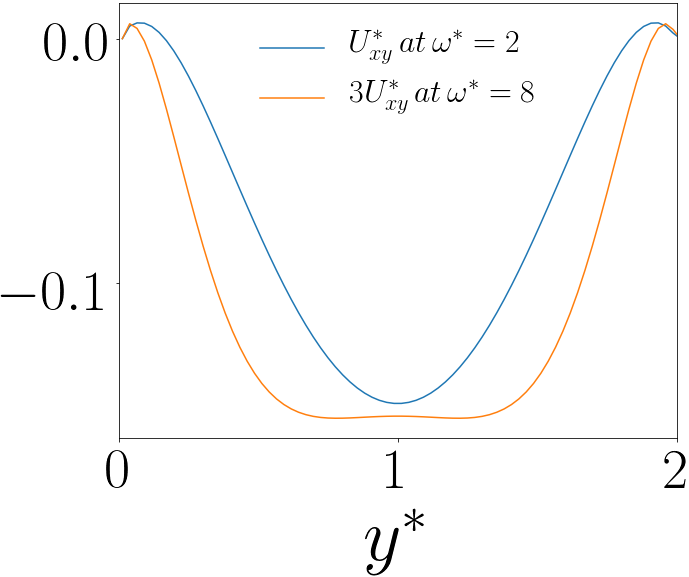}
  \caption{}
  \label{velocity comparison}
  \end{subfigure}
   \caption{Normalized flow direction averaged velocity profiles on the $xy$ and $xz$ planes at $Re=10$, $Ca=1.67$, and (a) $\omega^*=2$ $(Wo=3.15, St=3)$, (b) $\omega^*=8$ $(Wo=6.3, St=6)$, and (c) their shape comparison on the $xy$ plane with the same constant $z$}
     \label{velocity profiles}
\end{figure}

Stokes number in the oscillatory and pulsating flows is defined as $St=W\sqrt{\frac{\omega}{\nu}}$. \citeauthor{o1975pulsatile} \cite{o1975pulsatile} has solved these types of flows in rectangular channels analytically and quantified the shape of the velocity profile by calculating the ratio between the velocity at the center and its average across the cross-section. Based on the values in table 1 of this paper, the profile maintains its parabolic shape up to $St=3$ for a square channel. Above this approximate value, the profile starts to become more like a flat, plug-like profile \cite{o1975pulsatile}. Figure \ref{velocity profiles} illustrates the shapes of the dimensionless averaged velocity along the flow direction near the drop focal points at $St=3$ ($Wo=3.15$) and $St=6$ ($Wo=6.3$). These shapes are consistent with the findings of \cite{o1975pulsatile} and \cite{karbaschi2014high}. At $Wo=3.15$ ($\omega^*=2$) the velocity shape is still parabolic and is similar to those of the lower $Wo$ values. However, this shape changes to plug like at $Wo=6.3$ ($\omega^*=8$). The average of $V_{avg}$ at this $Wo$ is the lowest because it has the highest frequency among others. Therefore, the value of deformation force on average is very small according to equation \ref{deformation force}. However, due to the shape of the velocity profile, the relative flow velocity from the drop frame of reference is very small near the center and very large near the wall. Thus, there is a strong shear gradient force although the average of $V_{avg}$ is small. Consequently, the focal point at $Wo=6.3$ does not follow the trend observed for the lower $Wo$ numbers and is pushed towards the wall. 

Furthermore, by taking a closer look at Fig. \ref{velocity comparison}, we can see that the velocity has an opposite sign near the wall. Due to the present hysteresis in this type of flow, there is a lag in the response of fluid to the change of flow direction \cite{noguchi2010dynamic, chaudhury2016droplet}.

\begin{figure}[ht]
  \centering
  \includegraphics[width=3.3in]{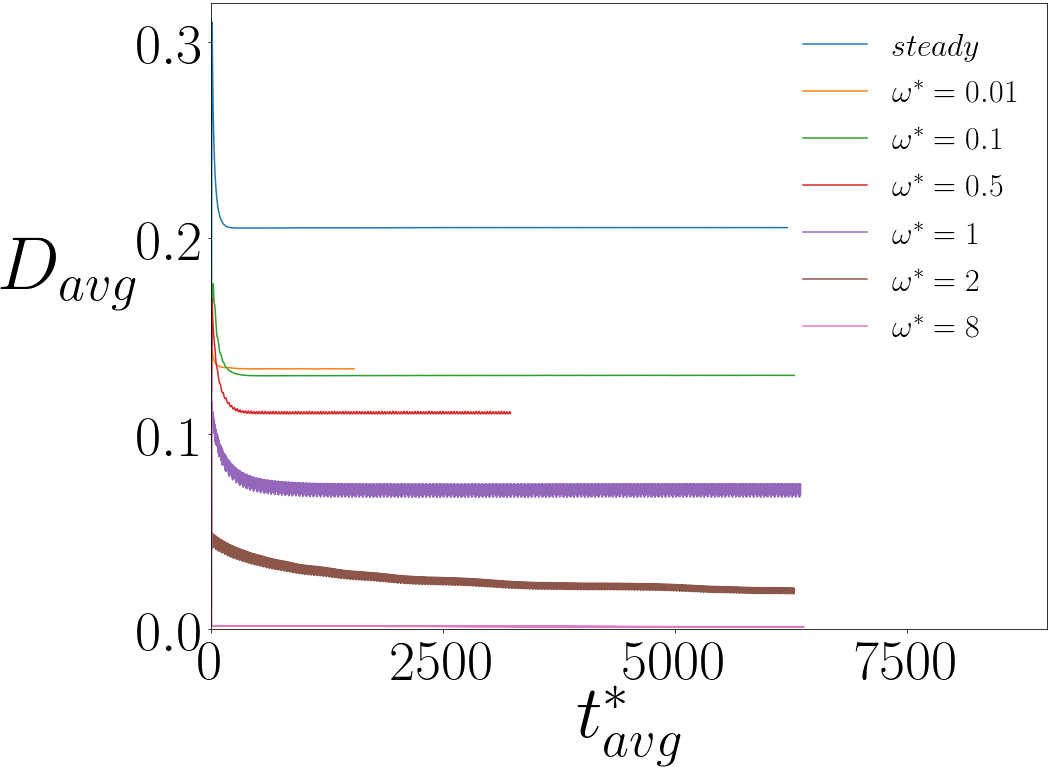}
   \caption{Averaged deformation parameter vs averaged dimensionless time at $Re=10$, $Ca=1$, and different frequencies}
   \label{deformation}
\end{figure}

The Taylor deformation parameter of the drop is defined as:
\begin{equation}
  D=\frac{L-B}{L+B}
\end{equation}
In which, $L$ is the principal major axis, and $B$ is the principal minor axis of an equivalent ellipsoidal particle. The parameter $D$ is oscillatory for all the flows in this study except for the steady one. This parameter is proportional to the shear rate, which depends on the flow velocity. As a result, the average of $D$ is lower at higher frequencies \cite{zhao2011dynamics}. This is reflected in Fig. \ref{deformation} by visualizing the average of $D$ in the corresponding periodic cycle as a function of time. This trend also implies that the amount of oscillations in the deformation is lower at higher frequencies since the minimum deformation in each cycle is zero. Moreover, it is apparent that the case with $\omega^*=8$ has a deformation of close to 0, which confirms that it has a very low $V_{avg}$. In fact, the droplet in this case travels about only $0.02W$ along the flow direction and remains almost spherical though the $Ca=1$ corresponds to a very deformable drop. Nevertheless, it migrates around $0.08W$ from the initial location to its focal point. 

Previous works have reported that drops in the flow regimes of high $Ca$ are elongated significantly leading to their break up \cite{lan2012numerical, pan2016motion, marson2018inertio}. This is the reason for the absence of any data for $Ca=10$ and $Re=10$ for steady and lower frequency flows in Fig. \ref{focal point}. The drop undergoes a very large deformation in these cases, which is not what it experiences at higher frequencies. A similar argument holds for $Re=37.8$ and $Ca=1.67$ since the drop is able to deform more easily at higher $Re$ \cite{inamuro2003lattice}.

\begin{figure}
\centering
  \begin{subfigure}[ht]{.485\textwidth}
  \centering
  \includegraphics[height=.225\textheight]{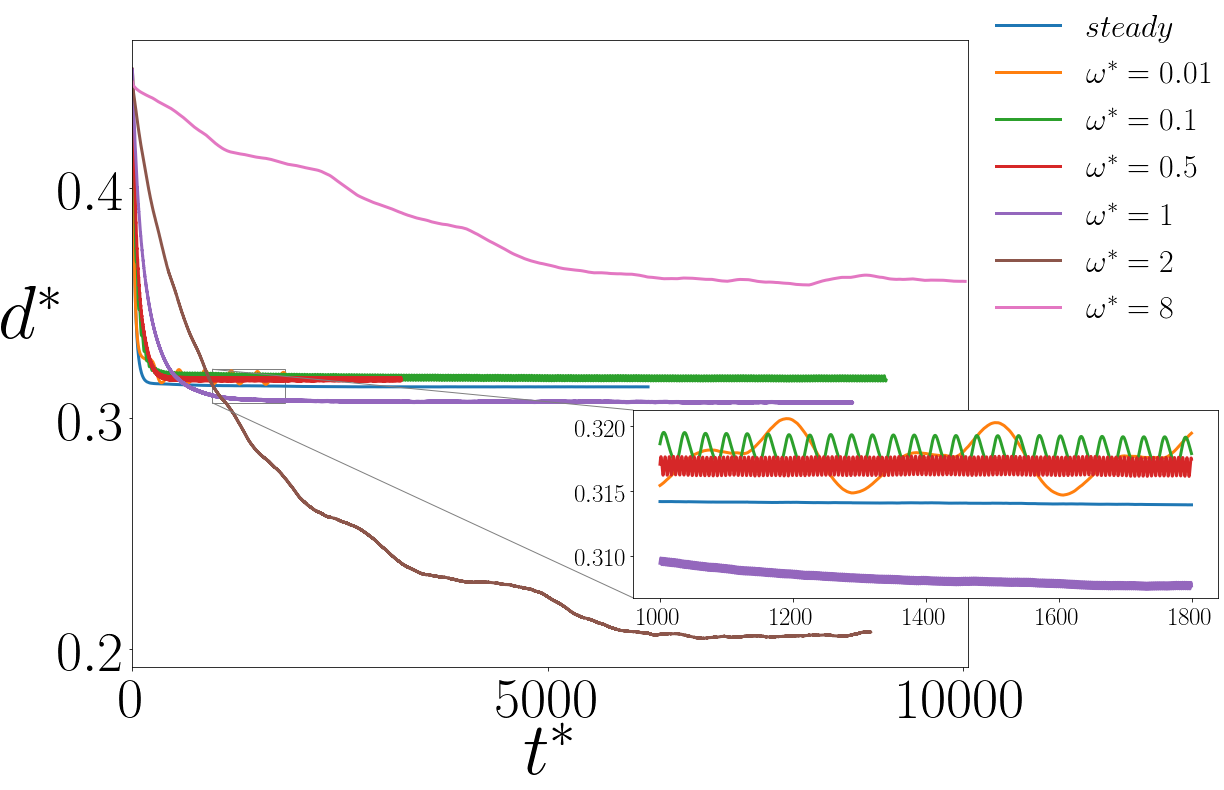}
  \caption{}
  \label{oscillation_Ca=1}
  \end{subfigure}
  ~
  \begin{subfigure}[ht]{.485\textwidth}
  \centering
  \includegraphics[height=.225\textheight]{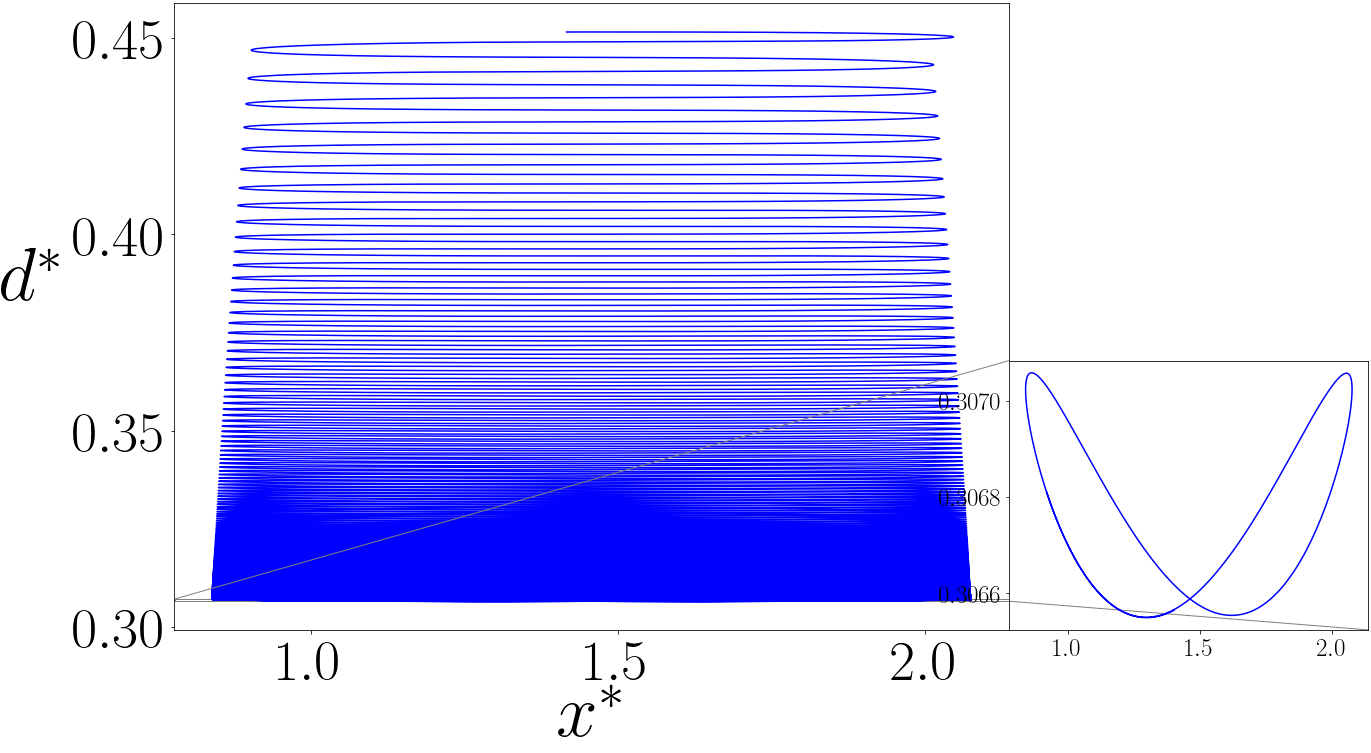}
  \caption{}
  \label{trajectory_vs_x_omega=1}
  \end{subfigure}
   ~
  \begin{subfigure}[ht]{.485\textwidth}
  \centering
  \includegraphics[height=.225\textheight]{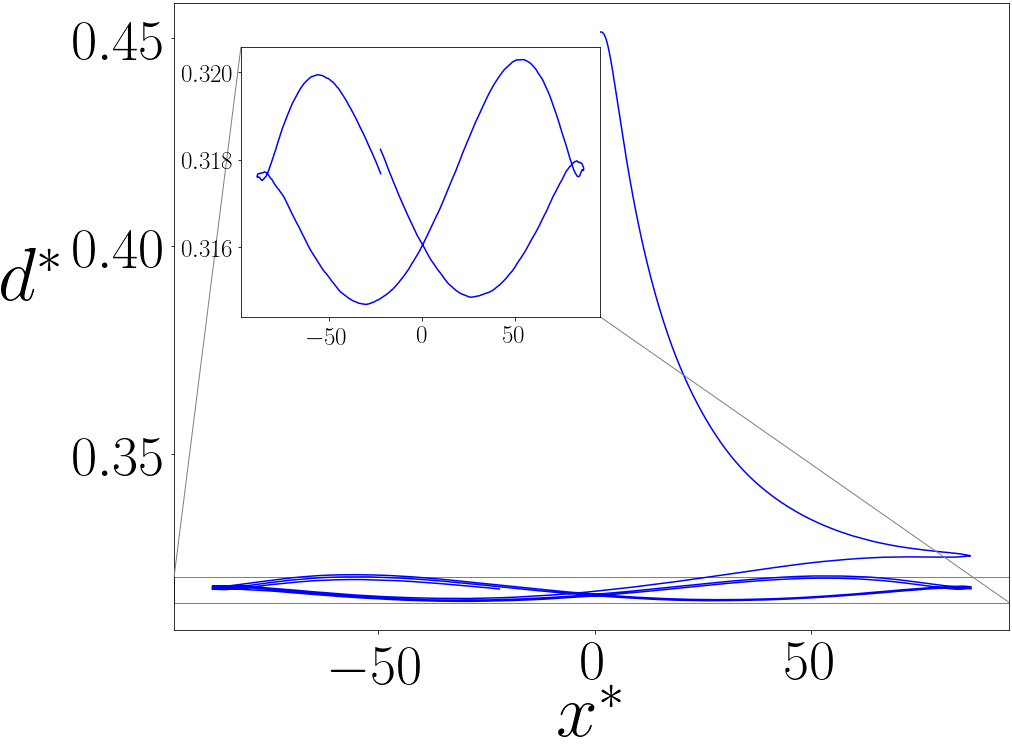}
  \caption{}
  \label{trajectory_vs_x_omega=0.01}
  \end{subfigure}
  ~
  \begin{subfigure}[ht]{.485\textwidth}
  \centering
  \includegraphics[height=.225\textheight]{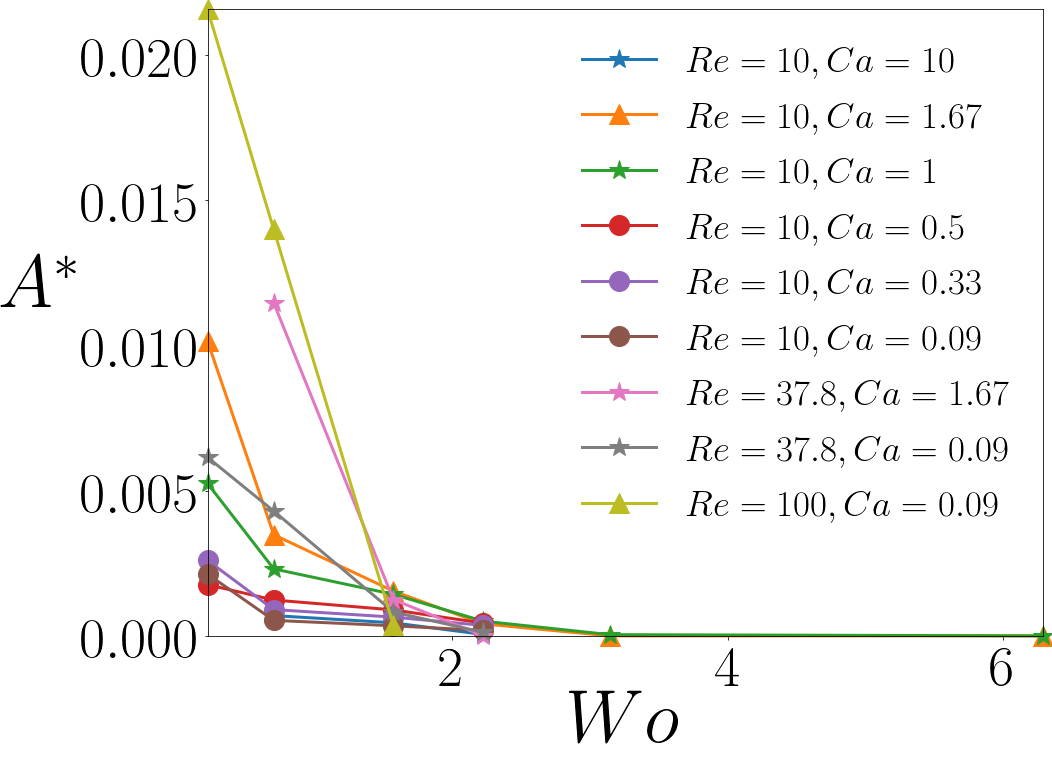}
  \caption{}
  \label{oscillation}
  \end{subfigure}
   \caption{(a) Migration patterns at $Re=10$, $Ca=1$, and various frequencies, (b) Drop trajectory versus the flow direction at $Re=10$, $Ca=1$, and $\omega^*=1$, and (c) $\omega^*=0.01$, and (d) Amplitude of oscillations around the equilibrium point after focusing for different cases}
     \label{oscillation_all}
\end{figure}

As it was discussed previously, the velocity in the presented type of flows is oscillatory. Consequently, all the active forces also experience fluctuations leading to oscillations in the trajectory of the drop except for the steady flows \cite{chaudhury2016droplet}. This can be seen in Fig. \ref{oscillation_all} where the whole lateral migration patterns at $Re=10$, $Ca=1$, and all frequencies, $\omega^*=1$, and $\omega^*=0.01$, showing the oscillations in the trajectories are illustrated in figures \ref{oscillation_Ca=1}, \ref{trajectory_vs_x_omega=1}, and \ref{trajectory_vs_x_omega=0.01}, respectively, and the amplitudes of oscillations after focusing ($A^*$) for different cases are depicted in Fig. \ref{oscillation}. In these figures, $d^*$ is the dimensionless, time-dependent distance of the drop from the channel center, and the insets of figures \ref{trajectory_vs_x_omega=1} and \ref{trajectory_vs_x_omega=0.01} show the trajectory at the last $2$ periodic cycles. The plots depicted in the latter $2$ figures qualitatively agree with those of a previous study and become more like a helical, spiral pathway as the frequency increases \cite{chaudhury2016droplet}. The minimum velocity in each periodic cycle is zero, occurring when the flow direction changes. The higher average velocity at lower frequencies implies a higher maximum velocity in the corresponding cycle. The higher the difference between the maximum and minimum velocities, the higher is the oscillations amplitude in the forces and in the trajectory. Therefore, similar to the aforementioned discussion of deformation oscillations, the parameter $A^*$ decreases as $\omega^*$ or $Wo$ increases (Fig. \ref{oscillation}). This is also the case for the oscillations amplitude along the flow direction (the comparison between figures \ref{trajectory_vs_x_omega=1} and \ref{trajectory_vs_x_omega=0.01}). The existence of these oscillations at all $Wo$ values is noteworthy \cite{chaudhury2016droplet}. It is vital to mention that all the values reported in Fig. \ref{focal point} are the averages of $d^*$ in the last periodic cycle.  

\begin{figure}[ht]
  \centering
  \includegraphics[width=3.3in]{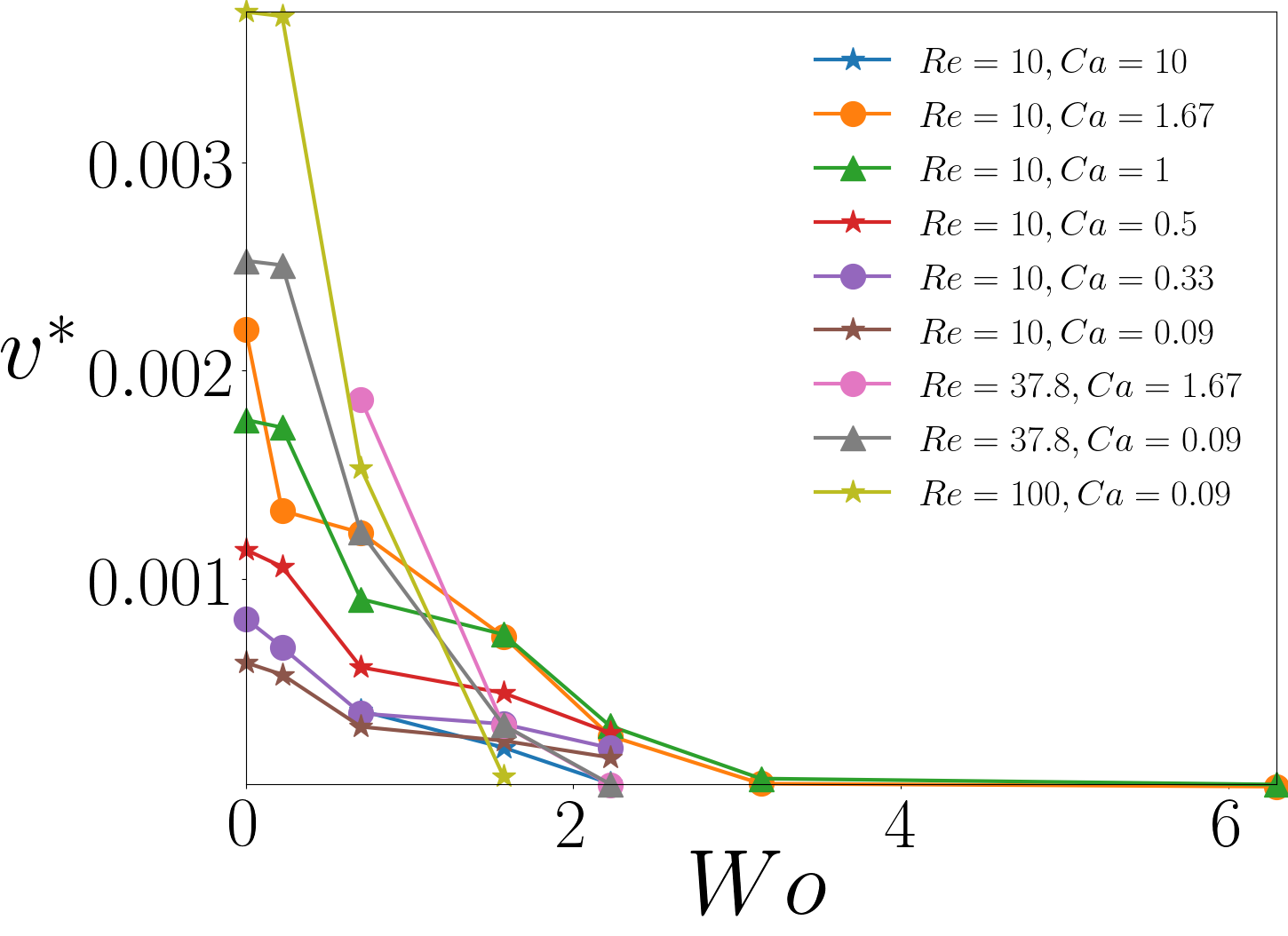}
   \caption{The average migration velocity vs $Wo$}
   \label{migration velocity}
\end{figure}

Focusing time can be considered as an important factor in the design and performance of the microfluidic system. Nevertheless, it strongly depends on where the particle is initially released. Therefore, the average migration velocity is a better parameter for a more meaningful comparison under different circumstances. The average migration velocity ($v^*$) for different cases is shown in Fig. \ref{migration velocity}. This velocity is computed by calculating the distance between the initial and equilibrium positions and dividing it by the focusing time. The corresponding focusing time is determined when the trajectory reaches within $0.015W$ of the focal point. This figure expresses that the average migration velocity decreases as $Wo$ increase. This pattern is also observed in the average velocity along the flow direction. Besides, the average migration velocity in the steady flows increases by increasing the $Ca$, which is in agreement with the findings of \citeauthor{alghalibi2019inertial} \cite{alghalibi2019inertial}.

\begin{figure}[ht]
  \centering
  \includegraphics[width=3.3in]{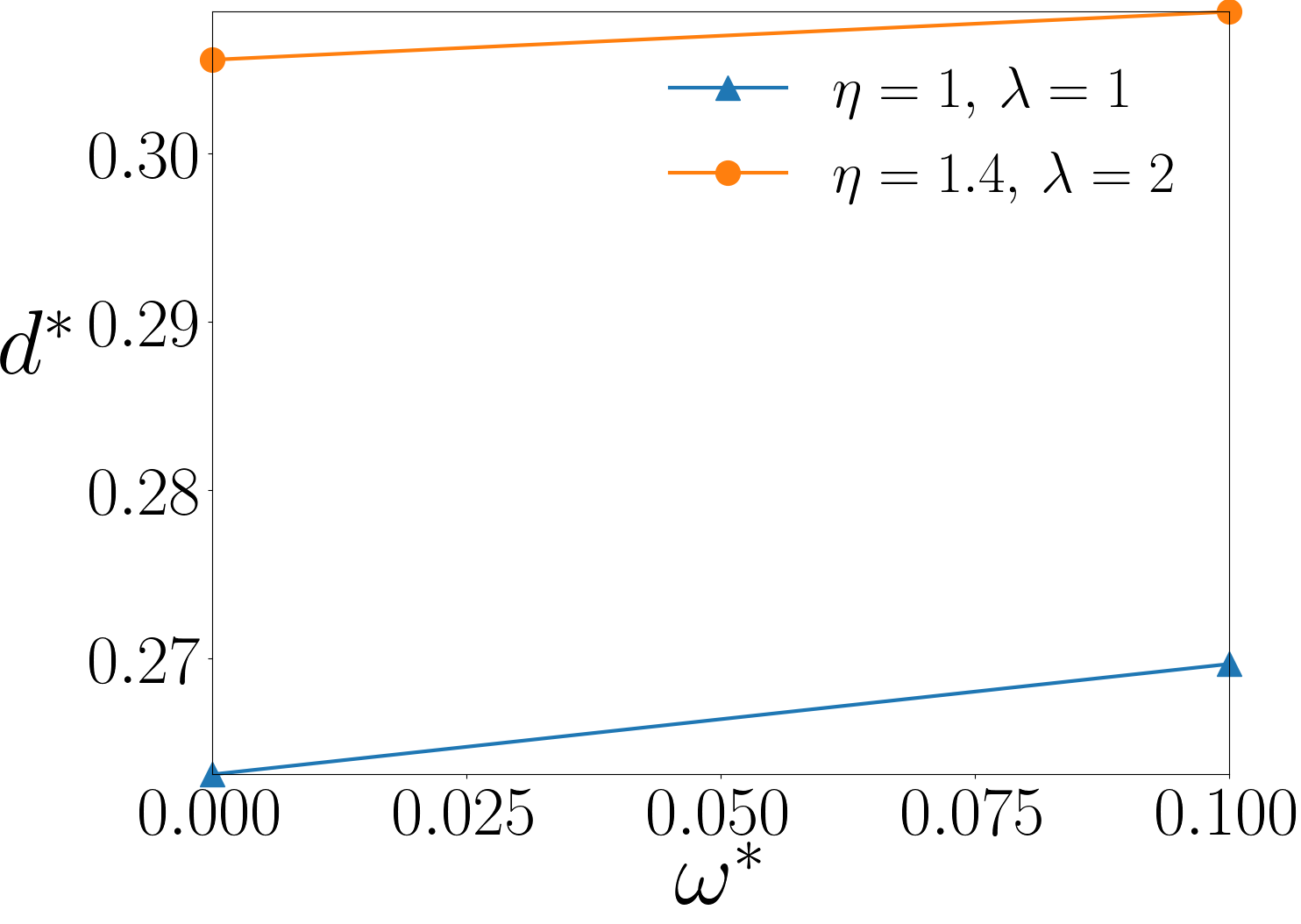}
   \caption{The effect of density and viscosity ratios on the distance of the drop focal point from the channel center at $Re=10$ and $Ca=1.67$}
   \label{viscosity ratio}
\end{figure}

Viscosity ratio in the range of $1\leq\lambda<13.2$ has a weak effect on the drop migration \cite{marson2018inertio, pan2016motion}. The effect of density ratio ($\eta$) on the drop focal point is even less \cite{mortazavi2000numerical}. Therefore, we have limited our study on the effect of these two parameters on the drop migration only to one case as shown in Fig. \ref{viscosity ratio}. The drop with a $\lambda$ higher than $1$ focuses closer to the wall for the $Ca$ we are studying here \cite{marson2018inertio, pan2016motion, mortazavi2000numerical}. We can also see that the change in the $\eta$ and $\lambda$ does not change the distance between the focal points in the steady and oscillatory flows significantly.
\subsection{Pulsating flow}
\begin{figure}
\centering
  \begin{subfigure}[ht]{.485\textwidth}
  \centering
  \includegraphics[height=.24\textheight]{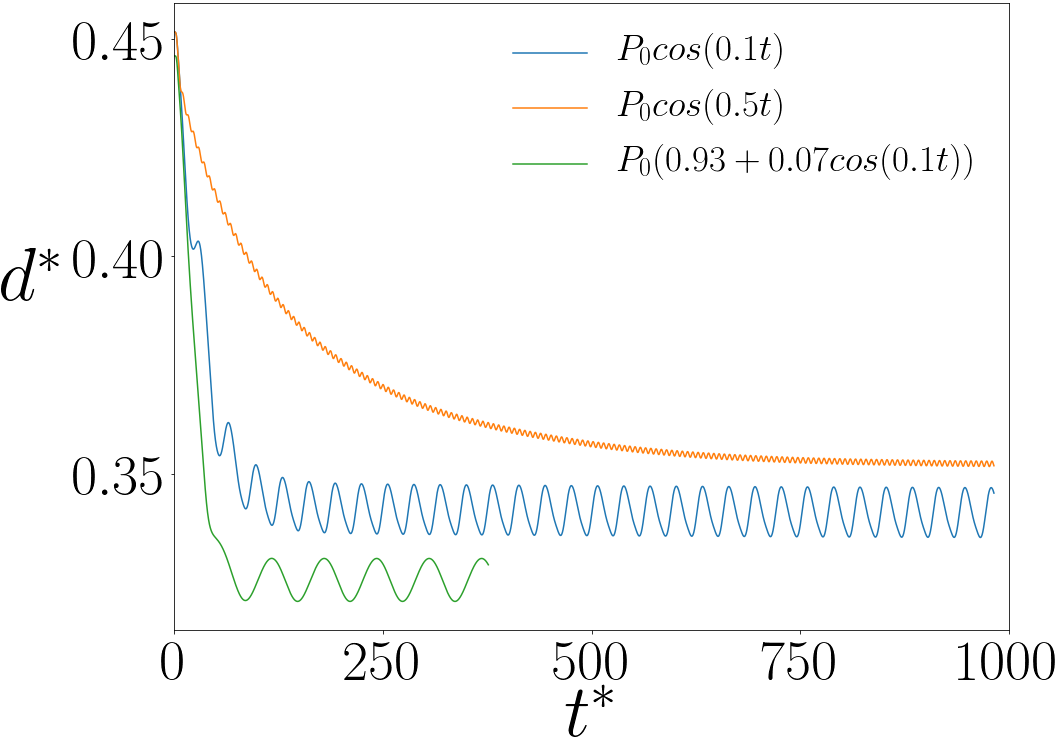}
  \caption{}
  \label{mean_Ca=1.67}
  \end{subfigure}
  ~
  \begin{subfigure}[ht]{.485\textwidth}
  \centering
  \includegraphics[height=.24\textheight]{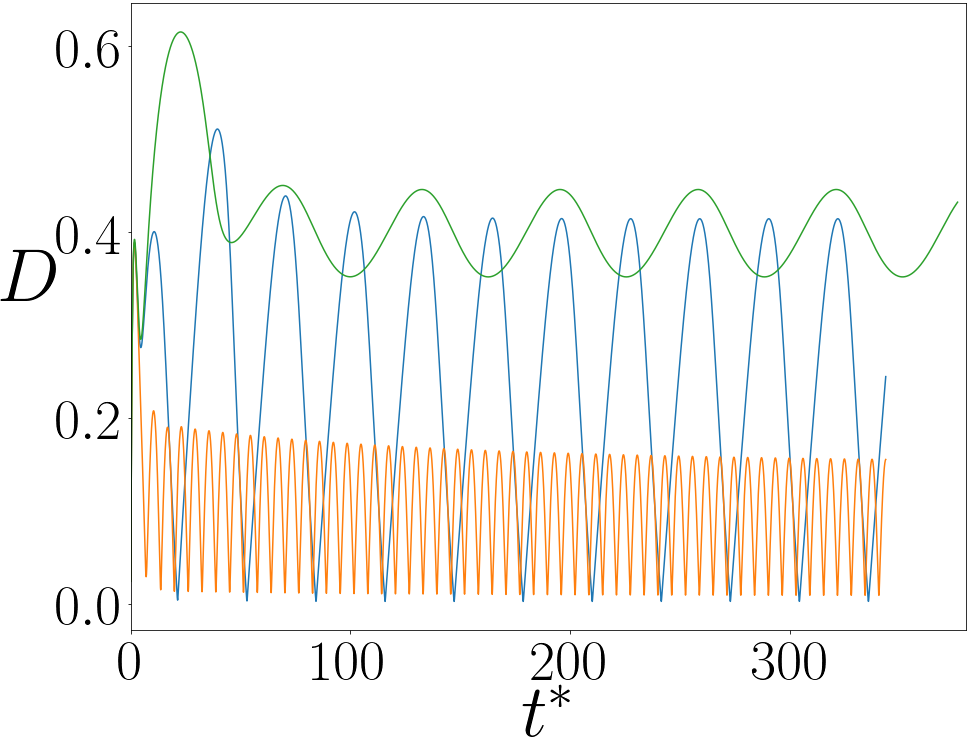}
  \caption{}
  \label{deformation_Ca=1.67}
  \end{subfigure}
  ~
  \begin{subfigure}[ht]{.485\textwidth}
  \centering
  \includegraphics[height=.24\textheight]{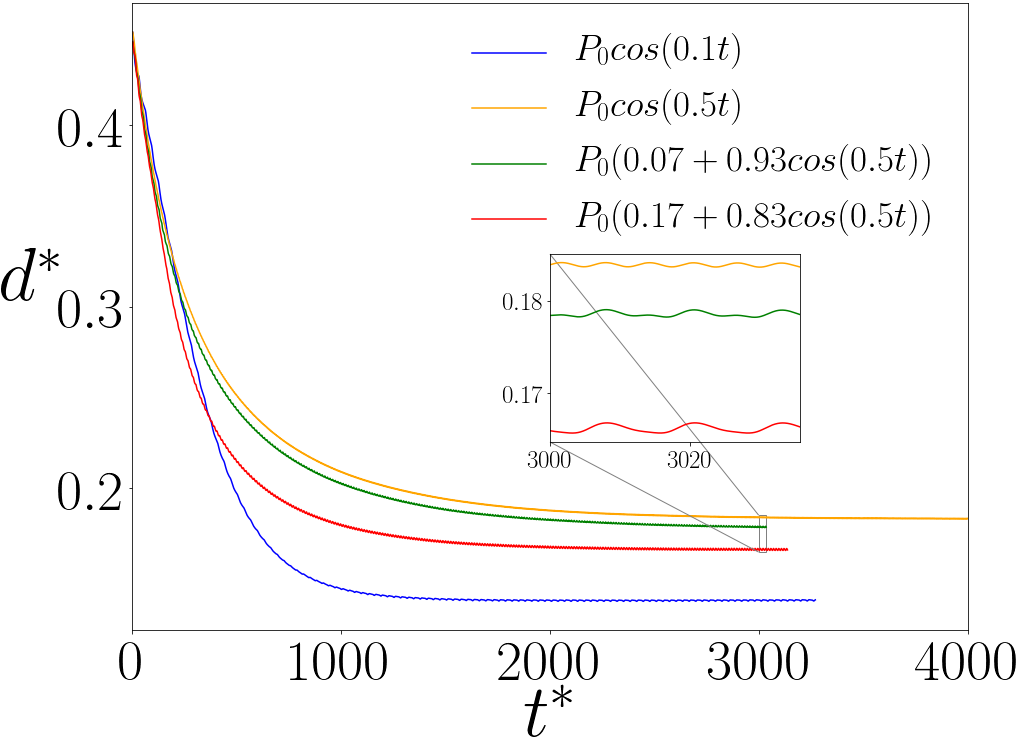}
  \caption{}
  \label{mean_Ca=10}
  \end{subfigure}
  ~
  \begin{subfigure}[ht]{.485\textwidth}
  \centering
  \includegraphics[height=.24\textheight]{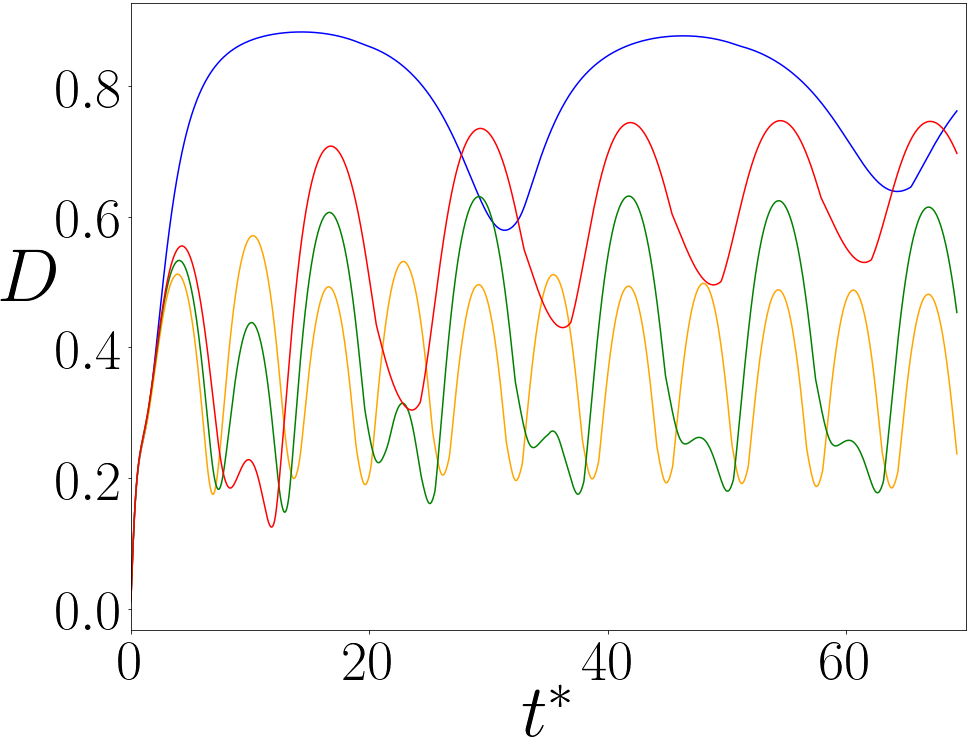}
  \caption{}
  \label{deformation_Ca=10}
  \end{subfigure}
   \caption{Evolution of the oscillatory and pulsating dynamics of the drop at $Re=37.8$ and $Ca=1.67$ by illustrating its (a) trajectory, (b) deformation, and at $Re=10$ and $Ca=10$ by visualizing the corresponding (c) trajectory, and (d) deformation}
     \label{pulsating}
\end{figure}

Changing the steady flow to oscillatory type helps achieve different focal points, which can have potential applications in cell sorting and separation as a great advantage. However, the oscillatory flow has a zero net throughput, which can be counted as a disadvantage considering many microfluidic applications depending on high-throughput systems. One solution to fix this issue is to combine both steady and oscillatory parts in the pressure gradient and make the flow regime to be pulsating \cite{dincau2020pulsatile}. A pulsating flow has the advantages of having non-zero net throughput as well as introducing a new equilibrium point for the drop. The latter occurs because the drop dynamics in the pulsating flow is very similar to the one in a pure oscillatory flow with an equivalent frequency between $0$ and that of the oscillatory portion of the pulsating pressure gradient. This claim is valid since the weights of steady and oscillatory portions add up to 1. We can see this equivalent frequency in Fig. \ref{mean_Ca=1.67} and the inset of Fig. \ref{mean_Ca=10} where the trajectories in the pulsating flows have frequencies of almost half of those of the oscillatory ones with the same frequencies. This feature also enables the existence of cases with high $Ca$ or $Re$ and low $\omega^*$ that are absent in the figures of the previous section. This is because the equivalent frequency of the pulsating flow makes the average deformation lower compared to the pure steady flow or oscillatory flow with a lower frequency. It is crucial to mention that the highest weights for the steady portions of the pulsating cases depicted in Fig. \ref{pulsating} denote flows in which the drop experiences the highest feasible deformation without breaking up or being significantly elongated. This has been done to show the highest possible changes in the focal points. 

In addition, the directionality of the focal points and averages of deformation values obey the expected trend at each combination of $Re$ and $Ca$ in Fig. \ref{pulsating}. For instance, the pulsating flow at $Re=37.8$ and $Ca=1.67$ has an equivalent frequency less than $0.1$. Therefore, we can see a focal point closer to the channel center and a higher average deformation according to Fig. \ref{mean_Ca=1.67} and Fig. \ref{deformation_Ca=1.67}, respectively. Similarly, we can observe a focal point closer to the center at $Re=10$ and $Ca=10$ for the pulsating case with the larger steady portion. This is because this case has an equivalent frequency between $0.1$ and $0.5$ and less than the one with the lower steady portion. Fig. \ref{mean_Ca=10} reflects this pattern. The deformation behavior in Fig. \ref{deformation_Ca=10} is also qualitatively similar to that shown in Fig. \ref{deformation_Ca=1.67}. It is also momentous to pay attention to the $0.012W$ difference between the focal points of the pulsating cases at $Re=10$ and $Ca=10$ although the difference between the weights of their steady portions is only $0.01$. 

Tables \ref{tab:$Re=37.8$ and $Ca=1.67$} and \ref{tab:$Re=10$ and $Ca=10$} quantify the average migration velocities for the flow regimes discussed above. The observed trend in these tables is consistent with the information provided in the discussion of Fig. \ref{migration velocity}. The pulsating case at $Re=37.8$ and $Ca=1.67$ has the lowest equivalent frequency. Hence, it has the highest average migration velocity among others. Similarly, the $v^*$ at $Re=10$ and $Ca=10$ decreases as the equivalent frequency increases.
\begin{table}
\begin{minipage}[h]{.473\textwidth}
   \centering
   \begin{tabular}{|P{4.12cm}|P{4.12cm}|}
     \hline
     \textbf{Pressure gradient form} & \textbf{Average migration velocity} \\ \hline
     $P_0(0.93+0.07cos(0.1t))$ & 0.00191\\ \hline
     $P_0cos(0.1t)$ & 0.00141 \\ \hline
      $P_0cos(0.5t)$ & 0.00012 \\ 
     \hline
   \end{tabular}
   \caption{Average migration velocities for different flow regimes at $Re=37.8$ and $Ca=1.67$}
   \label{tab:$Re=37.8$ and $Ca=1.67$}
\end{minipage}\qquad
\begin{minipage}[h]{.473\textwidth}
   \centering
   \begin{tabular}{|P{4.12cm}|P{4.12cm}|}
     \hline
      \textbf{Pressure gradient form} & \textbf{Average migration velocity} \\ \hline
     $P_0cos(0.1t)$ & 0.000389 \\ \hline
     $P_0(0.17+0.83cos(0.5t))$ & 0.000297 \\ \hline
     $P_0(0.07+0.93cos(0.5t))$ & 0.000213 \\ \hline
     $P_0cos(0.5t)$ & 0.000205 \\ 
        \hline
   \end{tabular}
   \caption{Average migration velocities for different flow regimes at $Re=10$ and $Ca=10$}
   \label{tab:$Re=10$ and $Ca=10$}
\end{minipage}
\end{table}

The effect of droplet size on its focal point in the oscillatory and pulsating flows at $Re=10$ and $Ca=10$ is summarized in Table \ref{tab:size}. We observe that reducing the drop size pushes its equilibrium location towards the wall \cite{pan2016motion, mortazavi2000numerical, di2009particle, wang2017analysis, bazaz2020computational}. Furthermore, this size reduction appears to enhance the change in the focal point at different values of equivalent frequency.
\begin{center}
\centering
    \begin{tabular}{|P{4.8cm}|P{3.5cm}|P{3.5cm}|}
      \hline
      \multirow{2}{*}{\textbf{Pressure gradient form}} & \multicolumn{2}{|c|}{\textbf{Equilibrium distance from center}} \tabularnewline
\cline{2-3} 
       & $\mathbf{\frac{a}{W}=0.3}$ & $\mathbf{\frac{a}{W}=0.2}$\tabularnewline
      \hline
      $P_0cos(0.1t)$ & 0.138 & 0.193\tabularnewline
      \hline
      $P_0(0.17+0.83cos(0.5t))$ & 0.166 & 0.241\tabularnewline
      \hline
      $P_0cos(0.5t)$ & 0.183 & 0.266\tabularnewline
      \hline
    \end{tabular}
    \captionof{table}{The effect of drop size on its focal point at $Re=10$ and $Ca=10$}
    \label{tab:size}
  \end{center}
\subsection{MFGP}
\begin{figure}[ht!]
  \centering
  \includegraphics[width=3.3in]{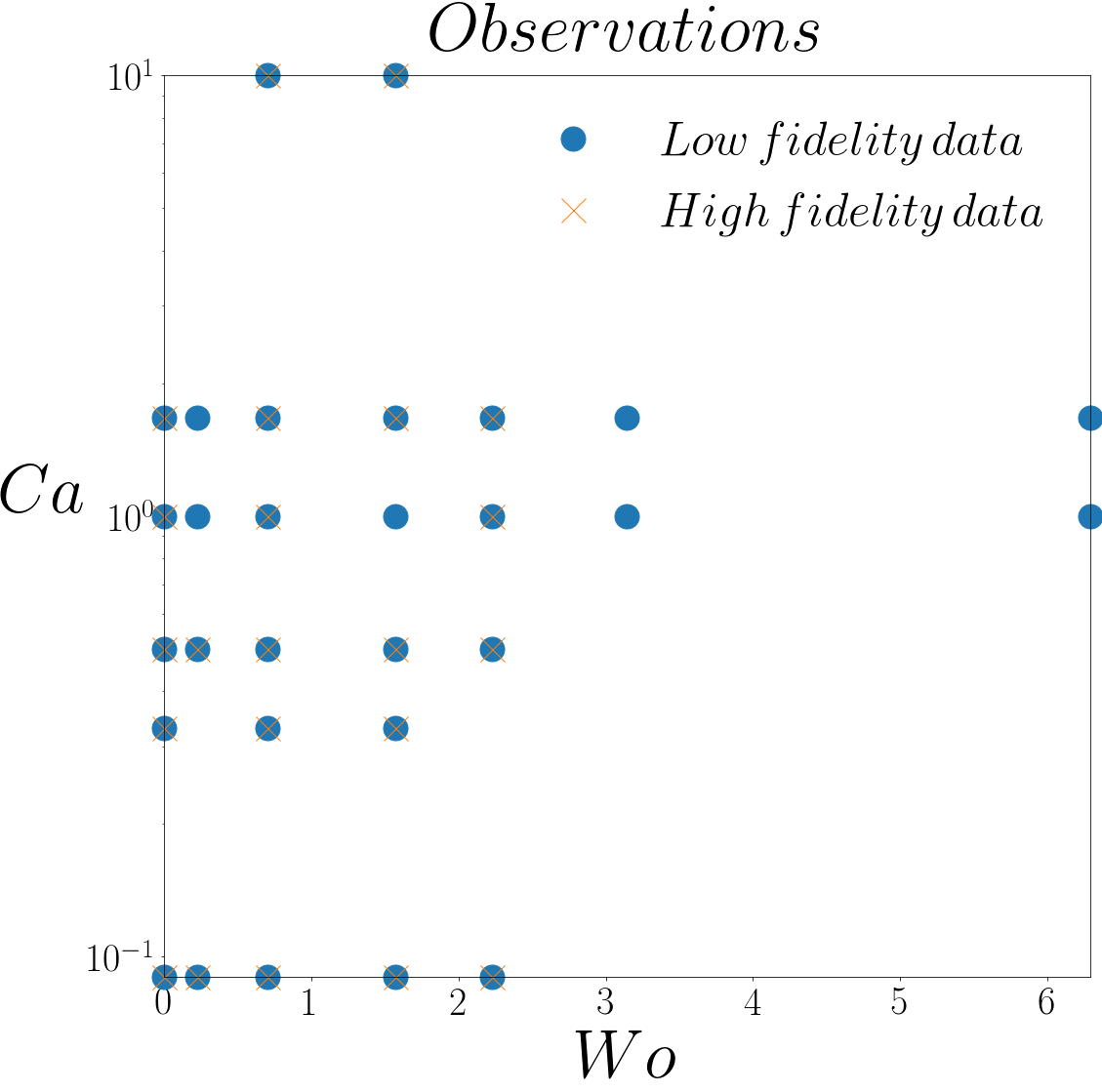}
   \caption{Observations of the high and low fidelity levels}
   \label{observations}
\end{figure}

\begin{figure}
\centering
  \begin{subfigure}[ht!]{.485\textwidth}
  \centering
  \includegraphics[height=.3\textheight]{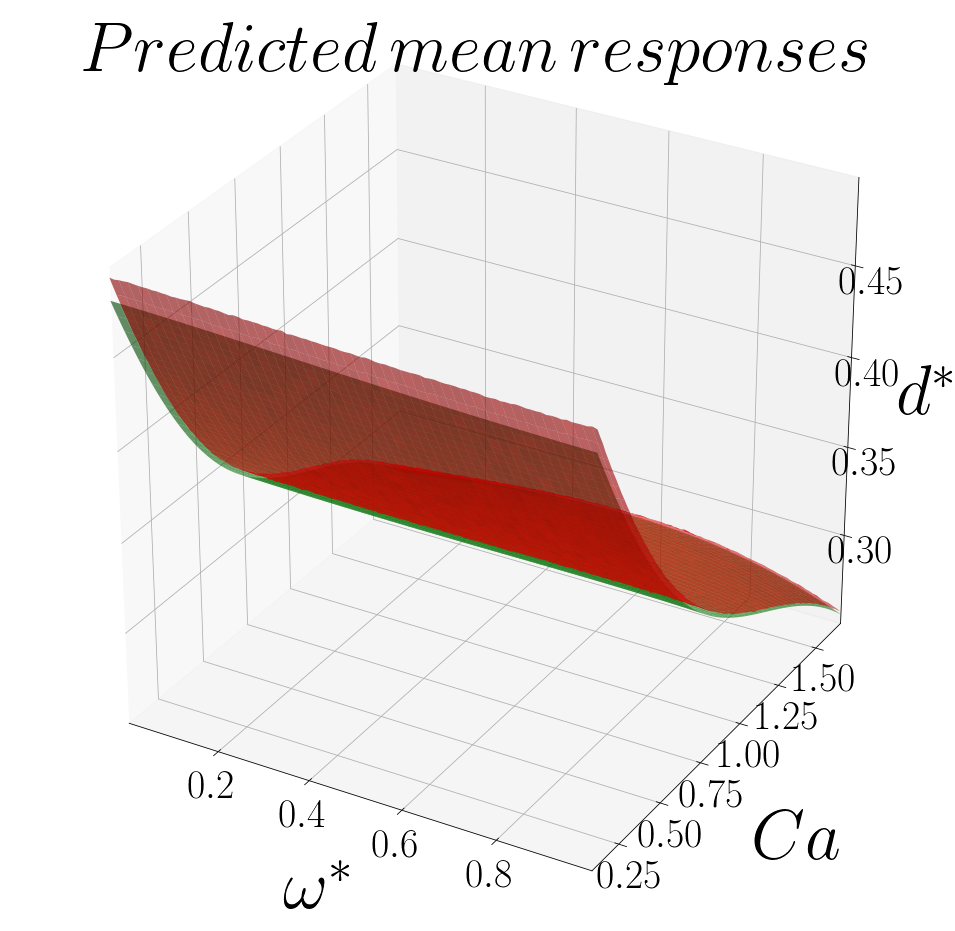}
  \caption{}
  \label{surface}
  \end{subfigure}
  ~
  \begin{subfigure}[ht!]{.485\textwidth}
  \centering
  \includegraphics[height=.3\textheight]{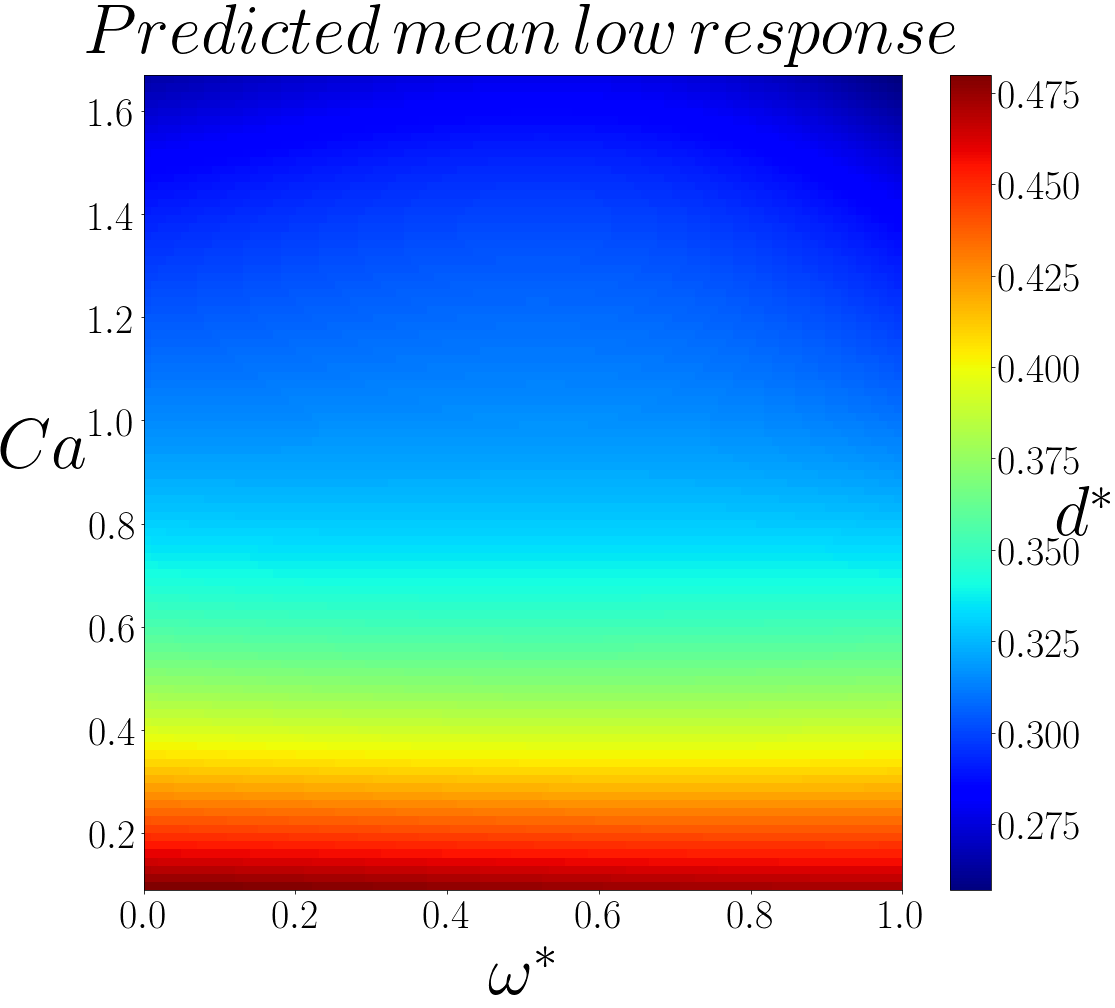}
  \caption{}
  \label{low}
  \end{subfigure}
  ~
  \begin{subfigure}[ht!]{.485\textwidth}
  \centering
  \includegraphics[height=.3\textheight]{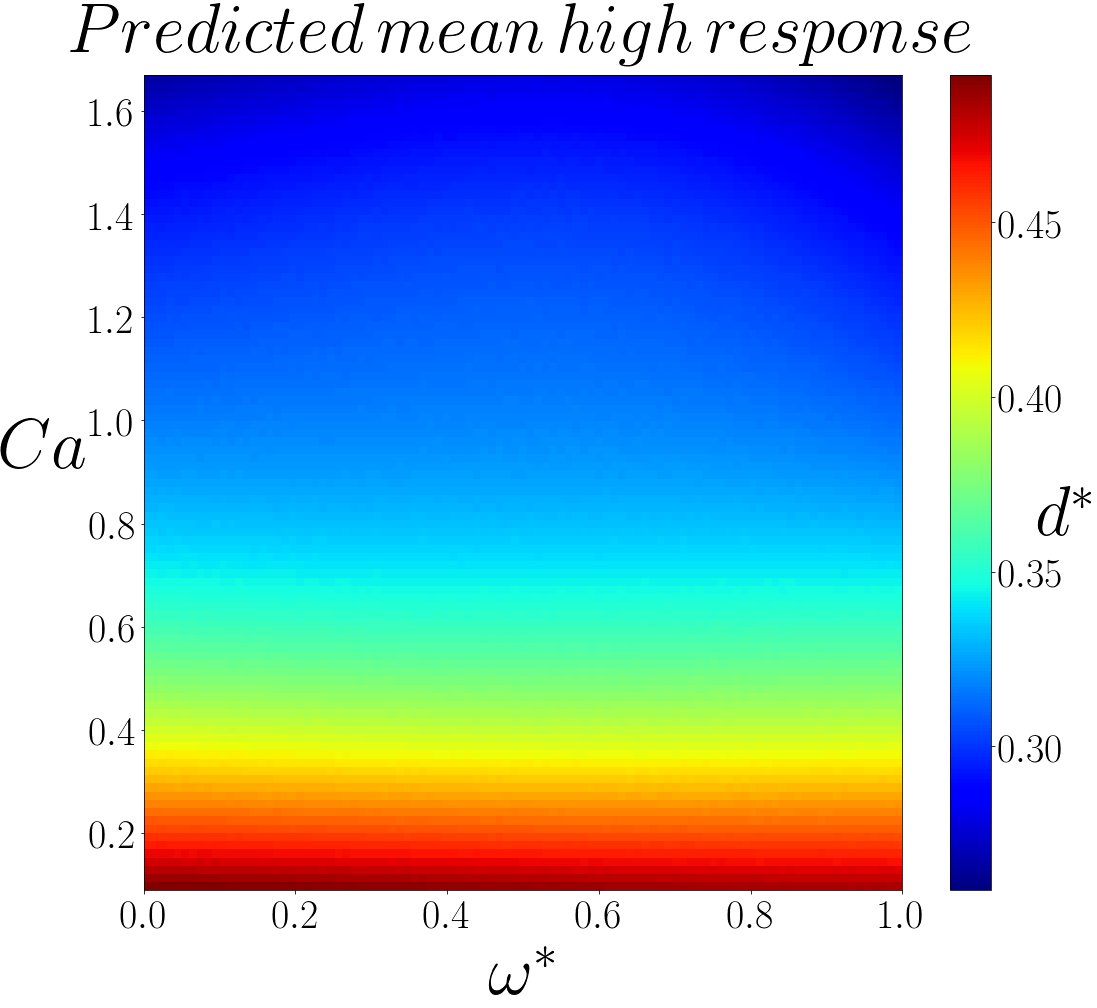}
  \caption{}
  \label{high}
  \end{subfigure}
  ~
  \begin{subfigure}[ht!]{.485\textwidth}
  \centering
  \includegraphics[height=.3\textheight]{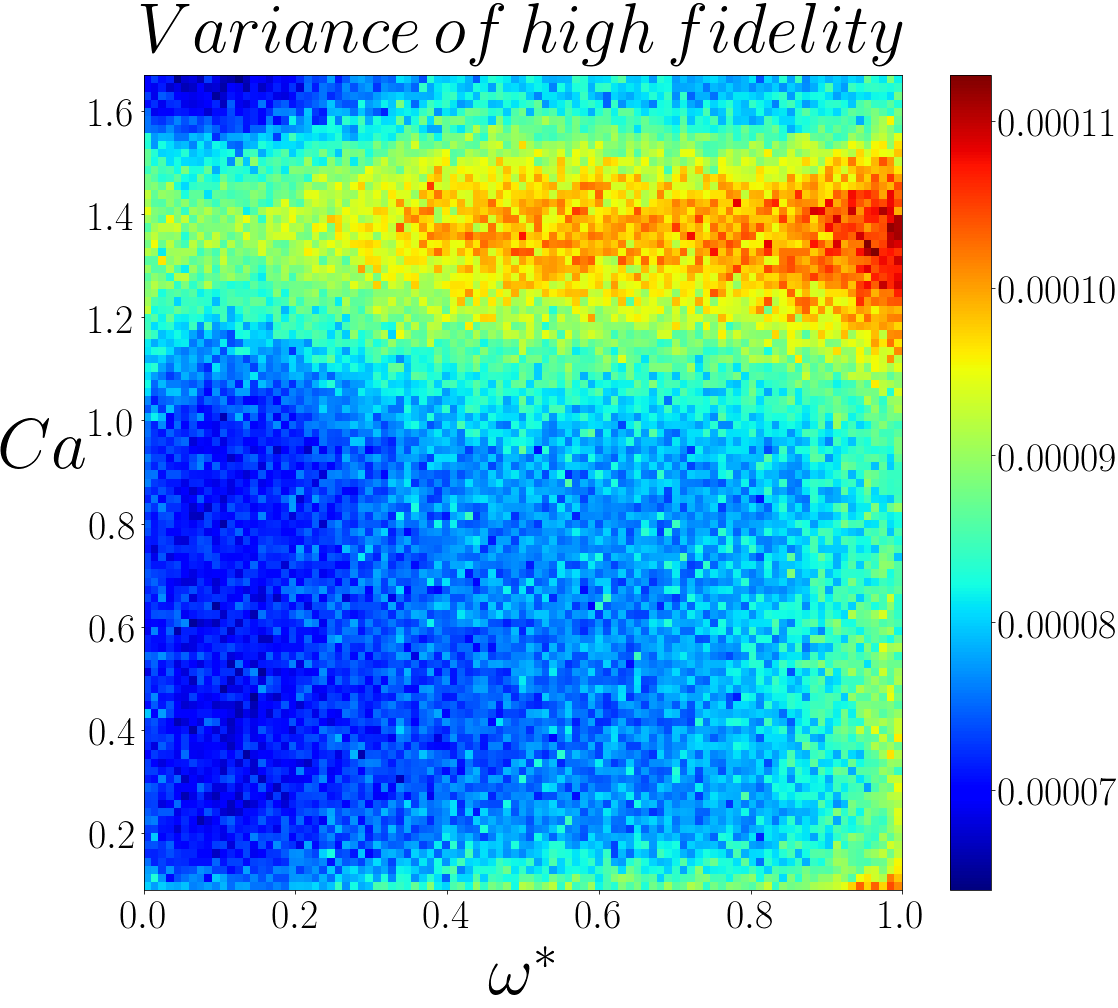}
  \caption{}
  \label{variance}
  \end{subfigure}
   \caption{Prediction of the MFGP algorithm by illustrating the (a) surface plot of the high-fidelity response colored by red and low-fidelity response colored by green, (b) mean response of the low-fidelity level, (c) mean response of the high-fidelity level, and (d) variance of the high-fidelity response}
     \label{predictions}
\end{figure}

In this section, we evaluate the MFGP performance by having a dataset consisting of 29 low-fidelity and 22 high-fidelity observations. Fig. \ref{observations} shows all these observations together. It can be seen that the required nested structure, as mentioned earlier, is satisfied. In other words, for any data point in the high-fidelity level, there is a corresponding point in the low-fidelity level. 
Fig. \ref{predictions} illustrates the predictions on the distance of the drop equilibrium position from the channel center over the range of 0 to 1 for the frequency and 0.09 to 1.67 for the Capillary number. The algorithm is trained over the entire available data. A few sanity checks have been done to ensure that these results make sense. First, Fig. \ref{surface} denotes that at any fixed value of the Capillary number, there is a global optimum frequency for which the equilibrium distance from the center (the $z$ axis) is an extremum. This is also previously observed in the simulations for the input values in the dataset (Fig. \ref{focal point}). Secondly, the predictions made by the high-fidelity response are slightly higher than those of the low-fidelity response, which is also compatible with the simulations outcomes. Lastly, we know from the underlying physics that in the steady flow (at a frequency of 0), the distance decreases by increasing the Capillary number \cite{pan2016motion, lan2012numerical}. This can be seen in both contour plots of the mean low and high responses (figures \ref{low} and \ref{high}) as well as being confirmed quantitatively in the code. Fig. \ref{variance} illustrates a relatively low variance for the high-fidelity response, being our main goal, in the entire domain. The red-colored region in this plot corresponds to places where the density of data points is less, and hence, the predicted outputs have more uncertainty. 

\begin{figure}
\centering
  \begin{subfigure}[ht!]{.485\textwidth}
  \centering
  \includegraphics[height=.28\textheight]{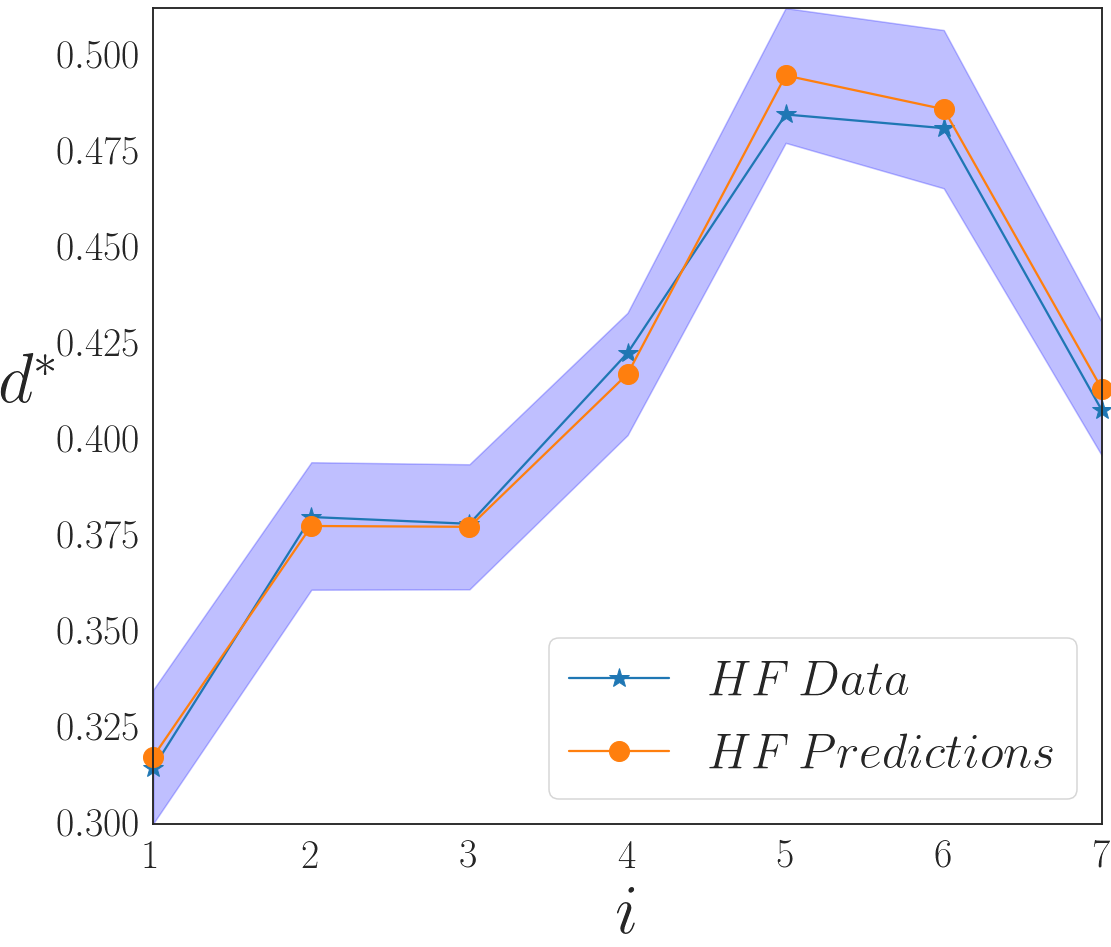}
  \caption{}
  \label{pred and obs}
  \end{subfigure}
  ~
  \begin{subfigure}[ht!]{.485\textwidth}
  \centering
  \includegraphics[height=.28\textheight]{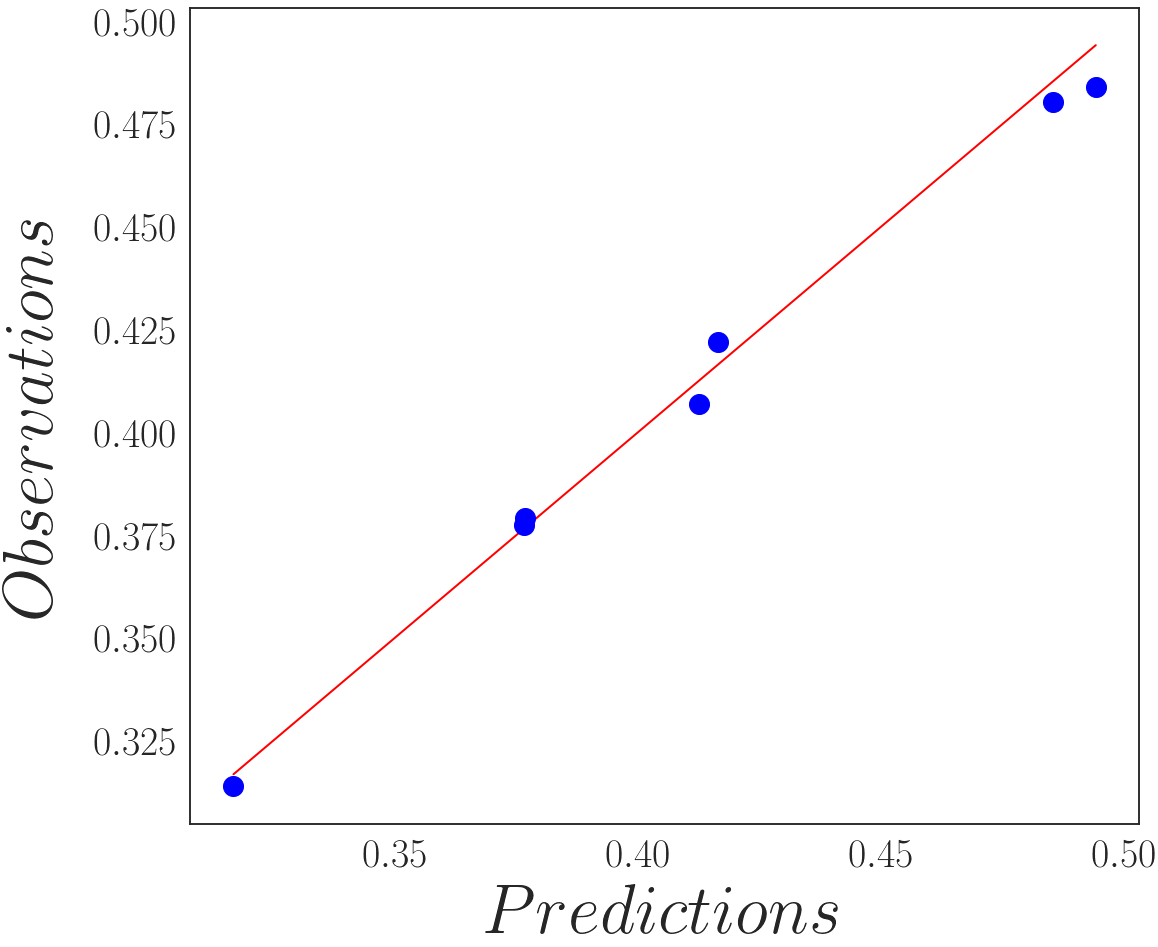}
  \caption{}
  \label{pred vs obs}
  \end{subfigure}
  ~
  \begin{subfigure}[ht!]{.485\textwidth}
  \centering
  \includegraphics[height=.28\textheight]{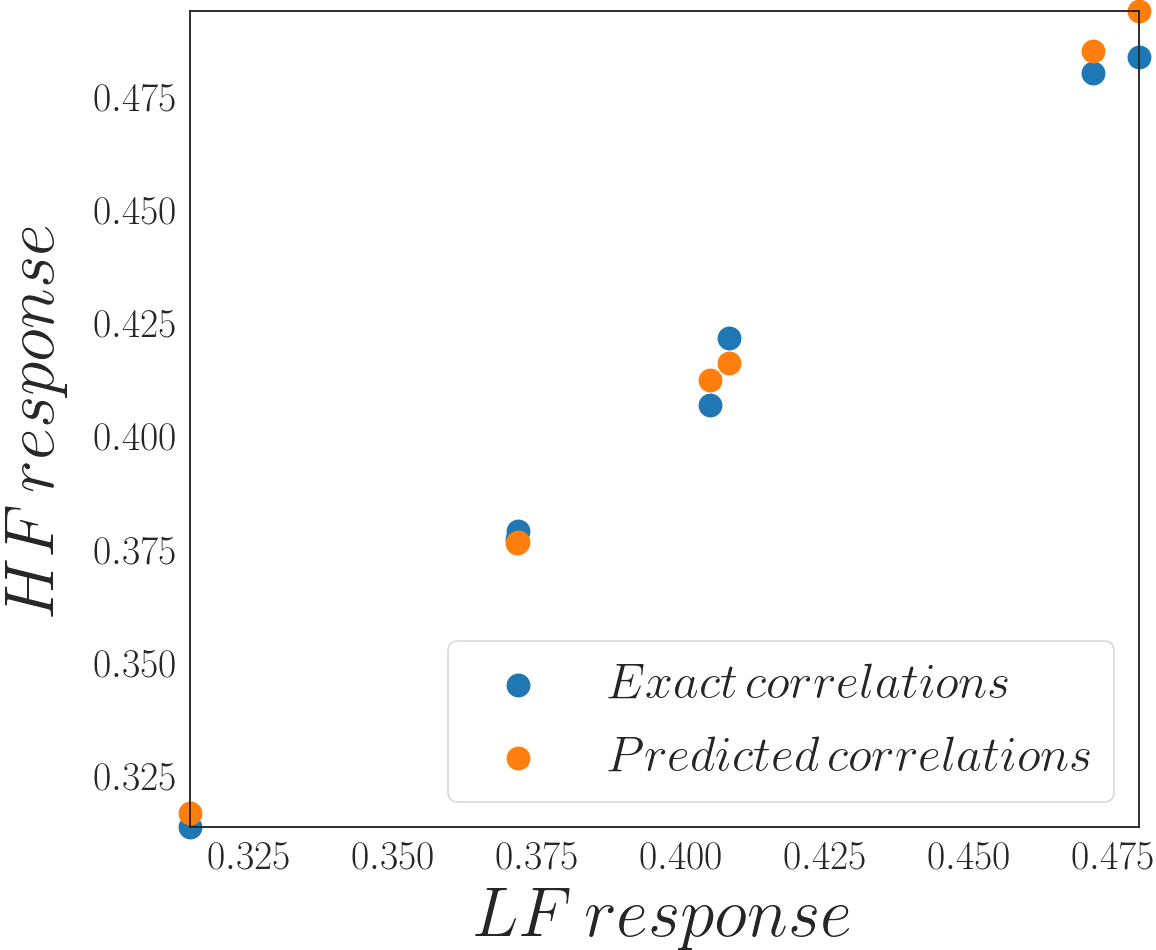}
  \caption{}
  \label{correlations}
  \end{subfigure}
  ~
   \caption{Evaluation of the MFGP performance on a random test set by illustrating the (a) observations and predictions with their uncertainties, (b) predictions versus observations, and (c) exact versus predicted correlations} 
     \label{test performance}
\end{figure}

Since there is no analytical solution to compare the predictions with, we decided to split the high-fidelity data into training and test sets in this section. This helps us evaluate the performance of the implemented MFGP. We assign 15 training and 7 test points at the high-fidelity level randomly. Then, we train the MFGP on the whole low-fidelity data (29 points) and only the training high-fidelity data, and evaluate the predictive responses on the high-fidelity test points, since those are our main targets. We do this entire procedure 500 times to eliminate the dependence of the results on the test points choice. This especially helps us examine the algorithm performance at the regions with less amount of data. After this bootstrapping, the average of the mean squared error (MSE) was 0.00015 and the average of the $R^2$ score was 0.9858. The successful reproduction of these results is also checked. Fig. \ref{test performance} shows this evaluation at the last (500th) test set. Fig. \ref{pred and obs} expresses the predictions of $d^*$ along with their 95\% credible intervals and denotes that the observed data lies within the shaded uncertainty bands. In this figure, the x axis label (i) denotes the index of each test point. Figure \ref{pred vs obs} visualizes the true or known value of $d^*$ versus its predicted value at each test point. This figure illustrates the strength of the model as the plotted points are very close to the line of $y=x$, which is an indication of the agreement between predictions and observations. The exact and predicted correlations between the high and low responses of $d^*$ are also very close to each other according to Fig. \ref{correlations}. All of these are evidence for the strong and successful performance of the implemented MFGP algorithm.
\section{Conclusions}
Determination and control of the particle's equilibrium position in the microchannels are extremely crucial as it can help in a variety of microfluidics applications. This importance, as well as the need to overcome the issue of designing impractically long channels to work with sub-micron particles, led us to do some simulations to capture the dynamics of a single droplet suspended in an oscillatory flow within the channel. The drawback of the zero net throughputs of the oscillatory flow has been addressed by modifying it further to become a pulsating flow. Both types of flows bring new equilibrium locations to the system. They also enable the presence of droplets at high $Ca$ or $Re$ that could break up in the steady or a very low-frequency regime. Moreover, fluctuations in the trajectory of the drop have been observed. It has been shown that the amplitude of these oscillations, the average of the oscillatory deformation, and the average migration velocity all decrease by increasing the frequency. The dependence of the drop focal point on the shape of the velocity profile has been investigated as well. It has been explored that this equilibrium position moves towards the wall in a plug-like profile, which is the case in very high $Wo$ numbers. Due to the significant cost of these simulations, a recursive version of the Multi Fidelity Gaussian processes has been used to replace the numerous high-fidelity simulations that cannot be afforded numerically. The MFGP algorithm is used to predict the equilibrium distance of the drop from the channel center for a given range of the interplaying input parameters, namely the Capillary number and frequency, assuming a constant Reynolds number. In addition, its performance was evaluated by randomly shuffling the high-fidelity data 500 times and assigning 31.8\% of it as the test set for an accurate quantitative comparison each time. The algorithm outputs high statistical scores, which is an indication of its reasonably accurate performance.

\bibliography{sorsamp}

\end{document}